\documentclass[
superscriptaddress,
preprint,
prd,tightenlines,nofootinbib,
eqsecnum]{revtex4-2}

\usepackage{amsfonts,amsmath,amssymb}
\usepackage[usenames,dvipsnames]{xcolor}
\usepackage{bm}
\usepackage{graphicx} 
\usepackage[colorlinks]{hyperref}
\usepackage{mathrsfs}
\usepackage{multirow}
\usepackage{tensor}
\usepackage{tabularx}
\newcolumntype{Y}{>{\centering\arraybackslash}X}
\AtBeginDocument{\usepackage{booktabs}}

\definecolor{darkgreen}{rgb}{0,0.5,0}

\hypersetup{
    bookmarks=true,         
    unicode=false,          
    pdftoolbar=true,        
    pdfmenubar=true,        
    pdffitwindow=false,     
    pdfstartview={FitH},    
    pdftitle={My title},    
    pdfauthor={Author},     
    pdfsubject={Subject},   
    pdfcreator={Creator},   
    pdfproducer={Producer}, 
    pdfkeywords={keyword1} {key2} {key3}, 
    pdfnewwindow=true,      
    colorlinks=true,       
    linkcolor=red,          
    citecolor=cyan,        
    filecolor=magenta,      
    urlcolor=darkgreen,           
    linktocpage=true
}

\newcommand{\di}{\mathrm{i}}

\newcommand{\ud}{\mathrm{d}}
\newcommand{\uD}{\mathrm{D}}

\newcommand{\calO}{\mathcal{O}}

\newcommand{\ph}[1]{\phantom{#1}}
\newcommand{\vph}[1]{\vphantom{#1}}
\newcommand{\phfrac}{\vphantom{\frac{1}{2}}}

\newcommand{\be}{\begin{equation}}
\newcommand{\ee}{\end{equation}}

\newcommand{\exch}{1\leftrightarrow 2}
\newcommand{\nn}{\nonumber}
\newcommand{\dd}{\mathrm{d}}

\usepackage{ulem}
\newcommand{\tmass}{M}
\newcommand{\ADM}{\mathcal{M}}

\allowdisplaybreaks

\usepackage{etoolbox}
\makeatletter
\patchcmd{\frontmatter@abstract@produce}
  {\vskip200\p@\@plus1fil
   \penalty-200\relax
   \vskip-200\p@\@plus-1fil}
\makeatother

\begin{document}

\title{Spin contributions to the gravitational-waveform modes for spin-aligned binaries at the 3.5PN order}

\author{Quentin \textsc{Henry}}\email{quentin.henry@aei.mpg.de}
\affiliation{Max Planck Institute for Gravitational Physics\\ (Albert Einstein Institute), D-14476 Potsdam, Germany}

\author{Sylvain \textsc{Marsat}}\email{sylvain.marsat@l2it.in2p3.fr}
\affiliation{Laboratoire des 2 Infinis - Toulouse (L2IT-IN2P3), Universit\'e de Toulouse, CNRS, UPS, F-31062 Toulouse Cedex 9, France}

\author{Mohammed \textsc{Khalil}}\email{mohammed.khalil@aei.mpg.de}
\affiliation{Max Planck Institute for Gravitational Physics\\ (Albert Einstein Institute), D-14476 Potsdam, Germany}

\date{\today}

\begin{abstract}
We complete the post-Newtonian (PN) prediction at the 3.5PN order for the spin contributions to the gravitational waveforms emitted by inspiraling compact binaries, in the case of quasi-circular, equatorial orbits, where both spins are aligned with the orbital angular momentum. Using results from the multipolar post-Minkowskian wave generation formalism, we extend previous works that derived the dynamics and gravitational-wave energy flux and phasing, by computing the full waveform decomposed in spin-weighted spherical harmonics. This new calculation requires the computation of multipolar moments of higher multipolar order, new quadratic-in-spin contributions to the hereditary tail terms entering at the 3.5PN order, as well as other non-linear interactions between moments. When specialized to the test-mass limit, our results are equivalent to those obtained in the literature for the waveform emitted by a test-mass in equatorial, circular orbits around a Kerr black hole. We also compute the factorized modes for use in effective-one-body waveform models, correcting the 2.5PN nonspinning and 3PN quadratic-in-spin terms in the (2,1) mode used in current models.
\end{abstract}


\maketitle


\section{Introduction}\label{sec:intro}

Since the first gravitational-wave (GW) detection in 2015 \cite{GW150914}, the LIGO-Virgo-KAGRA collaboration has observed over 90 GW signals from stellar-mass compact binary objects~\cite{LVCO1O2,catalogO3a,catalogO3b}. Future GW detectors, such as LISA \cite{LISA17} and the Einstein Telescope \cite{ET10}, will also widen the range in parameter space of detectable systems, including signals from extreme-mass-ratio inspirals and supermassive binary black holes, in addition to improving our knowledge of the deformability of neutron stars.

The use of accurate waveform templates for data analysis is crucial and requires constant improvement to match the increasing sensitivity of GW detectors. 
The post-Newtonian (PN) scheme is well suited to describe the inspiral of compact binaries, but its accuracy deteriorates in the strong-field regime.
The gravitational self-force (GSF) framework~\cite{Mino:1996nk,Quinn:1996am,Barack:1999wf,Detweiler:2002mi,Pound:2012nt,Barack:2018yvs,Pound:2021qin} is valid in the strong field, but for small-mass ratios.
However, by combining different analytical approximation methods and numerical-relativity results, the effective-one-body (EOB) formalism~\cite{BDEOB99,BuonD00} produces accurate waveforms over the entire parameter space.

Spin has a significant effect on the binary dynamics; thus, improving the spin description in waveform models is necessary to obtain accurate parameter estimations, which helps in improving our understanding of the properties of compact binaries and their astrophysical formation channels~\cite{LIGOScientific:2018jsj,LIGOScientific:2020kqk,LIGOScientific:2021psn}.
To model spin effects in general relativity, we use an effective field theory (EFT) approach, which constructs an effective action defined on the worldline of the bodies \cite{Mathisson37,Papa51spin,CPapa51,Tulc2,Dixon64,Dixon73,BOC75,BOC79,Dixon79,BIsrael75,
Po06,SP10,Harte12,LS15,LS15a,LS15c,M15}. In the conservative sector, many approaches were used to tackle the problem of the spin contributions in the equations of motion (EOM) which now reaches the 4.5PN accuracy \cite{KWW93,K95,TOO01,FBB06,SHS08a,SHS08b,SHS08c,PoR08a,PoR08b,PRR10,Po10,Le10so,Le10ss,Le12ss,
MBFB13,BMFB13,DJSspin,HSS10,HaS11,HaS11so,HaS11ss,HaSS13,BFH12,HaSS13,LS14,LS15a,LS15b,LS15c,M15,Marsat:2012fn,Levi:2016ofk,Levi:2019kgk,Vines:2016qwa,Antonelli:2020ybz,Antonelli:2020aeb}. On the other hand, the 3.5PN radiated flux and phase, including all spin contributions, were derived for quasi-circular orbits and non-precessing spins~\cite{KWW93,K95,MVGer05,BBF06,BBF11,BMB13,MBBB13,BFMP15,M15}. Recently, this computation has been pushed to 4PN by EFT methods \cite{Porto22Spins}. However, the polarizations for non-precessing spins are only known to 2.5PN order \cite{ABFO08,BFH12,PoRR12}. In the case of precessing spins, which complicate the computations, the waveform amplitude is known to 1.5PN order, and the phase to 2PN order~\cite{ABFO08}.

In the present article, we compute the spin effects in the full waveform up to the 3.5PN order. The dynamics and phase evolution are already known to that order, and the remaining piece is the computation of the amplitude of the waveform, decomposed as a sum of spin-weighted spherical harmonics. These harmonics complete our knowledge of the post-Newtonian waveform, and they are a crucial ingredient of waveform models such as the EOB approach~\cite{BuonD99,BDEOB99,Damour:2007xr,Damour:2007yf,Pan11,DIN09,Taracchini:2012ig,Bohe:2016gbl,Nagar:2016ayt,Nagar:2018zoe,Nagar:2019wds,Messina:2018ghh,Cotesta:2018fcv}.

At the 3.5PN order, three interactions arise: the spin-orbit (SO), the spin-induced quadrupole and the spin-induced octupole interactions. We use the post-Newtonian-multipolar-post-Minkowskian (PN-MPM) formalism \cite{BlanchetLR}, where the observables of the system are parametrized in terms of the so-called radiative multipole moments, defined in the radiative zone. The PN-MPM formalism allows linking these moments to another set of moments called the source multipole moments, which are defined over the source of the system and are linked to the metric and stress-energy tensor. The PN approximation is an expansion in the dimensionless quantity $v^2/c^2 \sim G\tmass/c^2r \ll 1$, where $v$ is the relative velocity and $\tmass$ is the total mass of the binary. To make PN expansions clear, we write $1/c$ explicitly in this paper, and rescale the physical spin variable $S_{\rm physical}$ as
\begin{equation}\label{eq:rescale}
S=c\, S_{\rm physical}=Gm^2\chi \,,
\end{equation} 
where $\chi$ is the dimensionless spin, whose value is one for an extremal Kerr black hole.\\

The paper is organized as follows. In Sec.~\ref{sec:summaryformalism}, we give some definitions and conventions, and recall the relation between the spherical modes and the radiative multipole moments. Sec.~\ref{sec:computations} contains the technical aspects regarding the computation of the radiative multipole moments. First in Sec.~\ref{subsec:spinsummary}, we recall some aspects of the effective action describing spinning effects up to the 3.5PN order. In Sec.~\ref{subsec:tmunu}, we give the expression of the stress-energy tensor as well as its 3+1 decomposition. Sec.~\ref{subsec:general def source multipole} gives the general definition of the source multipole moments. In Sec.~\ref{subsec:summarymetric} we give the PN metric and compute the potentials that parametrize it. Sec.~\ref{subsec:integration source multipole} is dedicated to the integration of the source multipole moments using the potentials and the decomposed stress-energy tensor. In Sec.~\ref{subsec:CoMcirc}, we express the source moments in the center of mass (CM) in the case of aligned spins and quasi-circular orbits. Finally, for this technical part, in Sec.~\ref{subsec:non-linear}, we derive the radiative multipole moments by computing the non-linear interactions in the GW field. In Sec.~\ref{sec:results}, we give the results for the spinning contributions to the amplitude modes, written in the conventional way in Sec.~\ref{subsec:hlm} and factorized conveniently for EOB usage in Sec.~\ref{subsec:eob}. Appendices \ref{app:sourcedensities} and \ref{app:eob} contain the lengthy expressions for the source densities and the factorized modes.
We also provide our results for the waveform modes as Mathematica files in the Supplemental Material~\cite{SuppMaterial}.


\section{Summary of the formalism}\label{sec:summaryformalism}
\subsection{Definitions and conventions}\label{subsec:notations}

In this paper, we compute the amplitude modes of the GW emitted by a binary system of spinning compact objects in the inspiral phase. We restrict ourselves to the non-precessing case which implies that, as for the non-spinning case, the motion of the system remains planar. Let us define the radiative coordinate system $X^\mu=(cT,\bm{X})$, in which $R=|\bm{X}|$ is the distance of the observer from the source and $\bm{N}=\bm{X}/R$ is the direction of propagation of the GW. We denote retarded time $T_R= T- R/c$, where $c$ is the speed of light in vacuum. We define the spatial unitary basis $(\bm{e}_X,\bm{e}_Y,\bm{e}_Z)$ such that the orbital plane lies within the $(\bm{e}_X,\bm{e}_Y)$ plane and $\bm{N}$ in the $(\bm{e}_Y,\bm{e}_Z)$ plane. The two GW polarizations $h_+$ and $h_\times$, defined in Eq.~\eqref{eq:hpluscross}, propagate in the plane orthogonal to the direction of propagation $\bm{N}$. It is convenient to introduce the orthonormal triad $(\bm{P},\bm{Q},\bm{N})$ by choosing $\bm{P}=\bm{e}_X$ and $\bm{Q}=\bm{N}\times\bm{P}$.\\

We use the following conventions henceforth: $\calO(n)$ means $\calO(1/c^{n})$, i.e.\ represents a contribution of the order $(n/2)$PN at least. Greek indices denote spacetime coordinates, i.e.\ $\mu = 0,1,2,3$, while Latin indices are used for spatial coordinates, i.e.\ $i=1,2,3$. We use the multi-index notation $L=i_1\dots i_\ell$. Symmetrization and anti-symmetrization are represented by, respectively, parenthesis and brackets around indices while the symmetric trace-free (STF) projection is denoted by $\langle\rangle$. We adopt the signature $(-,+,+,+)$ and keep explicit both Newton's constant $G$ and the speed of light $c$, unless explicitly specified. Finally the covariant derivative along the worldline is written as $\uD/(c\,\uD \tau) = u^{\mu}\nabla_{\mu}$, where $u^{\mu}$ is the four-velocity of the particle. 

The problem is parametrized using the following notations: we use the coordinate-time $t$ parametrization. The quantities $m_A$, $\bm{y}_A$, $\bm{v}_A=\dd \bm{y}_A/\dd t$ and $\bm{a}_A=\dd^2 \bm{y}_A/\dd t^2$ refer respectively to the mass, position, velocity and acceleration of body $A=1,2$. The notation SO refers to the spin-orbit interaction, SS to the quadratic-in-spin interaction and finally SSS to the cubic-in-spin terms.

\subsection{Spherical harmonics decomposition}

The Einstein field equations can be exactly written, by imposing the harmonic gauge condition $\partial_\nu h^{\mu\nu}=0$, as
\begin{equation}
\Box h^{\mu\nu}=\frac{16\pi G}{c^4}\tau^{\mu\nu},
\end{equation}
where $\Box$ is the flat d'Alembertian operator defined with respect to the inverse Minkowski metric $\eta^{\mu\nu}$, $h^{\mu\nu}$ is the deviation to the gothic metric $h^{\mu\nu}=\sqrt{-g}g^{\mu\nu}-\eta^{\mu\nu}$, $g=\text{det}(g_{\mu\nu})$ is the determinant of the metric. Finally, $\tau^{\mu\nu}$ is the stress-energy pseudo tensor 
\begin{equation}\label{eq:taumunu}
\tau^{\mu\nu} = \vert g\vert T^{\mu\nu}+\frac{c^4}{16\pi G}\,\Lambda^{\mu\nu}\,,
\end{equation}
where $T^{\mu\nu}$ is the stress-energy tensor and $\Lambda^{\mu\nu}$ is a function of derivatives of at least quadratic terms in the perturbed metric $h^{\mu\nu}$. Its expression is given in Eq.~(24) of Ref.~\cite{BlanchetLR}.\\

In the coordinate system $X^\mu=(cT,\bm{X})$, the transverse-traceless (TT) projection $h_{ij}^\text{TT}$ of the gravitational field $h^{\mu\nu}$ can be, at leading order in $1/R$, uniquely decomposed in terms of a set of STF multipole moments $U_L$ and $V_L$, called radiative multipole moments, as \cite{Th80}
\begin{align}\label{eq:hij}
h_{ij}^\text{TT} &= \frac{4G}{c^2R} \,\mathcal{P}_{ijkl} (\bm{N}) \sum^{+\infty}_{\ell=2}\frac{1}{c^\ell \ell !} \left\{ N_{L-2} \,U_{klL-2}(T_R) - \frac{2\ell}{c(\ell+1)} \,N_{aL-2} \,\varepsilon_{ab(k} \,V_{l)bL-2}(T_R)\right\} \,,
\end{align}
where we introduced the TT projection operator $\mathcal{P}_{ijkl} = \mathcal{P}_{i(k}\mathcal{P}_{l)j}-\frac{1}{2}\mathcal{P}_{ij}\mathcal{P}_{kl}$, with $\mathcal{P}_{ij}=\delta_{ij}-N_iN_j$ the projector orthogonal to the unit direction $\bm{N}$. We can define the usual polarization waveforms  in the orthonormal triad $(\bm{P},\bm{Q},\bm{N})$ as
\begin{subequations}\label{eq:hpluscross}
\begin{align}
h_+ &= \frac{1}{2}(P_i P_j - Q_i Q_j)h_{ij}^\text{TT},\\
h_\times &= \frac{1}{2}(P_i Q_j + Q_i P_j)h_{ij}^\text{TT}.
\end{align}
\end{subequations}
We can then decompose the quantity $h_+ -\di h_\times$ in a spin-weighted spherical harmonics basis of weight -2 \cite{K07}
\begin{equation}
h\equiv h_+ -\di h_\times = \sum_{l=0}^\infty \sum_{m=-\ell}^{\ell} h_{\ell m} Y^{\ell m}_{-2}(\Theta,\Phi),
\end{equation}
where the two angles $(\Theta,\Phi)$ characterize the direction of propagation $\bm{N}$ in the coordinate system $X^\mu$. In this paper, we follow the conventions of Ref.~\cite{BFIS08}. Notably, the explicit expression of the spin-weighted spherical harmonics are given in Eqs.~(2.4) of that reference. The amplitude modes $h_{\ell m}$ are then linked to the radiative moments $U_L$ and $V_L$ through
\begin{equation}\label{eq:hlm}
h_{\ell m} = -\frac{G}{\sqrt{2} R c^{\ell+2}}\left( U^{\ell m} -\frac{\di}{c} V^{\ell m} \right).
\end{equation}
where $U^{\ell m}$ and $V^{\ell m}$ read\footnote{Note that the choice of definitions in Ref.~\cite{K07,Pan11} on $h_{\ell m}$ differs from ours by a global minus sign for each $\ell$ and $m$ due to a different definition of the vector basis $(\bm{P},\bm{Q},\bm{N})$.}
\begin{subequations}\label{eq:UlmVlm}
\begin{align}
U^{\ell m} &= \frac{4}{\ell!}\,\sqrt{\frac{(\ell+1)(\ell+2)}{2\ell(\ell-1)}}\,\alpha_L^{\ell m}\,U_L\,,\\
V^{\ell m} &= -\frac{8}{\ell!}\,\sqrt{\frac{\ell(\ell+2)}{2(\ell+1)(\ell-1)}}\,\alpha_L^{\ell m}\,V_L\,.
\end{align}
\end{subequations}
The STF tensorial coefficient $\alpha_L^{\ell m} \equiv \int \dd \Omega\,\hat{N}_L\,\overline{Y}^{\,\ell m}$ is defined from the ordinary spherical harmonics $Y^{\ell m}$ (or in fact its complex conjugate $\overline{Y}^{\,\ell m}$) and its explicit expression is given in Eq.~(4.7) of Ref.~\cite{Jij3PN}.

The main result of this paper is the spin contributions in $h$ up to the 3.5PN order for a non-precessing, quasi-circular motion. In order to derive them, we have to compute the spin part of the radiative moments to consistent order. Table \ref{tab:cycles} shows at which order the different multipoles need to be computed for each spin interaction. The SO terms in the radiated flux and GW phase have been derived in the PN-MPM formalism up to 4PN while the SS contributions are known to 3.5PN~\cite{BMB13,MBBB13,BFMP15,M15}, however the radiative moments are required to a higher order for the modes than for the flux. This can be understood by comparing the expression~\eqref{eq:hij} of the full waveform, with a $1/c$ increment for each multipolar order, to the expression of the energy flux emitted in gravitational waves (e.g. (4.2) in Ref.~\cite{BFMP15}) as sum of squares of radiative moments with a $1/c^{2}$ increment per multipolar order.
\begin{table*}[h]
\begin{center}
\caption{
Leading and required (relative to the leading) orders of the spin contributions in the radiative moments for the full gravitational waveform at 3.5PN. The order corresponds to the power in $1/c$ and not the PN order.}
\label{tab:cycles}
\begin{tabularx}{0.95\textwidth}{cc *{9}{Y}}
\toprule
  & & \multicolumn{3}{c}{Leading order} & \multicolumn{3}{c}{Required relative order}\\
\cmidrule(lr){3-5} \cmidrule(l){6-8}
  $\ell$ & Moment & SO & SS & SSS & SO & SS & SSS \\
\midrule
  2 & $U_{ij}$          & 3 & 4 & 7 & 4 & 3 & 0 \\
  3 & $U_{ijk}$         & 3 & 4 & 7 & 3 & 2 & - \\
  4 & $U_{ijkl}$        & 3 & 4 & 7 & 2 & 1 & - \\
  5 & $U_{ijklm}$       & 3 & 4 & 7 & 1 & 0 & - \\
  6 & $U_{ijklmp}$      & 3 & 4 & 7 & 0 & - & - \\
  \hline
  2 & $V_{ij}$          & 1 & 4 & 5 & 5 & 2 & 1 \\
  3 & $V_{ijk}$         & 1 & 4 & 5 & 4 & 1 & 0 \\
  4 & $V_{ijkl}$        & 1 & 4 & 5 & 3 & 0 & - \\
  5 & $V_{ijklm}$       & 1 & 4 & 5 & 2 & - & - \\
  6 & $V_{ijklmp}$      & 1 & 4 & 5 & 1 & - & - \\
  7 & $V_{ijklmpq}$     & 1 & 4 & 5 & 0 & - & - \\
\bottomrule
\end{tabularx}
\end{center}
\end{table*} 


\section{Computation of the radiative multipole moments}\label{sec:computations}

In this paper, we use the PN-MPM formalism, see Ref.~\cite{BlanchetLR} for a review. Within this approach, the procedure to derive the radiative multipole moments is well-known. For completeness, we give a summarized version of the method, recalling the different steps: i) we first compute the stress-energy tensor from the effective skeletonized action modeling spinning particles in Sec.~\ref{subsec:spinsummary}; ii) in Sec.~\ref{subsec:tmunu}, we decompose the spatial and temporal indices of the stress-energy tensor to compute the source densities; iii) in Sec.~\ref{subsec:summarymetric}, we compute the PN metric parametrized by a set of potentials; iv) in Sec.~\ref{subsec:integration source multipole}, we derive the source multipole moments defined in Sec.~\ref{subsec:general def source multipole}; v) we then express the source moments in the CM frame for aligned spins and quasi-circular orbits in Sec.~\ref{subsec:CoMcirc}; vi) finally, we compute the non-linear effects to deduce the radiative multipole moments in Sec.~\ref{subsec:non-linear}.

\subsection{Effective skeletonized action of spinning particles}\label{subsec:spinsummary}

In this section, we start by recalling the method of the effective action approach to the spin-induced effects. We follow here the presentation of  \cite{BFMP15, M15}, but note that a more general formalism in the language of effective field theories can be found in Ref.~\cite{LS15a}. Here, we drop the particle's label $A$ and set $c=1$ for more clarity.

The matter action is constructed on the worldline of the individual body $z^\mu$. The affine parameter along this worldline is called $\tau$ and will be associated after variation of the action to the proper time of the body. We define the usual 4-velocity of the particle $u^\mu=\dd z^\mu/\dd\tau$ such that $u^{\mu}u_{\mu}=-1$. Attaching an orthonormal tetrad $\epsilon_{A}^{\phantom{A}\mu}$ to the moving body, we also introduce antisymmetric rotation coefficients $\Omega^{\mu\nu} = \epsilon^{A\mu} D \epsilon_{A}^{\phantom{A}\nu} /d\tau$ representing the rotational degrees of freedom. The matter action then reads symbolically
\begin{equation}
S_\text{M} = \int \dd^4 x \sqrt{-g} \int \dd \tau \,L_\text{M} [u^\mu,\Omega^{\mu\nu},g_{\mu\nu},R_{\mu\nu\rho\sigma},\nabla_\lambda R_{\mu\nu\rho\sigma}]\frac{\delta^{(4)}(x^\mu-z^\mu(\tau))}{\sqrt{-g}},
\end{equation}
where the matter Lagrangian includes finite-size effects in the form of couplings to the Riemann tensor and its derivative, necessary to represent spin-induced effects up to the cubic order in spin. We also introduce the following quantities 
\begin{align}\label{eq:dixonmoments}
p_\mu &\equiv\frac{\partial L}{\partial u^\mu}, \qquad\qquad S_{\mu\nu}\equiv2\frac{\partial L}{\partial \Omega^{\mu\nu}},\nonumber \\
J^{\mu\nu\rho\sigma}&\equiv-6\frac{\partial L}{\partial R_{\mu\nu\rho\sigma}}, \qquad J^{\lambda\mu\nu\rho\sigma}\equiv-12\frac{\partial L}{\partial \nabla_\lambda R_{\mu\nu\rho\sigma}},
\end{align}
that represent the couplings entering the action. The quantities $p^{\mu}$ and $S^{\mu\nu}$ are the linear momentum and the antisymmetric spin tensor. The moments $J^{\mu\nu\rho\sigma}$ and $J^{\lambda\mu\nu\rho\sigma}$ are Dixon-type moments \cite{Dixon64,Dixon73,Dixon79} representing the quadrupole and octupole, assumed to be purely spin-induced in our case (no other finite-size effects such as tidal effects). The general form of the stress-energy tensor will be expressed in terms of these quantities in the following section.

Varying the action with respect to the worldline and rotational degrees of freedom (together with a scalarity condition for the Lagrangian, see (2.8) in Ref.~\cite{M15}) yields evolution equations for the 4-momentum as well as the spin tensor:
\begin{subequations} \label{eq:evolution}
\begin{align}
	\frac{\uD p_{\mu}}{\uD\tau} &= -\frac{1}{2}R_{\mu\nu\rho\sigma}u^{\nu}S^{\rho\sigma} - \frac{1}{6} J^{\lambda\nu\rho\sigma} \nabla_{\mu} R_{\lambda\nu\rho\sigma} - \frac{1}{12} J^{\tau\lambda\nu\rho\sigma}\nabla_{\mu}\nabla_{\tau}R_{\lambda\nu\rho\sigma}+ \calO(S^4)\,, \label{eq:dpdt} \\
  \frac{\uD S^{\mu\nu}}{\uD \tau} &= 2p^{[\mu}u^{\nu]} + \frac{4}{3} R^{[\mu}_{\ph{\mu}\lambda\rho\sigma}J^{\nu]\lambda\rho\sigma} + \frac{2}{3}\nabla^{\lambda}R^{[\mu}_{\ph{\mu}\tau\rho\sigma} J_{\lambda}^{\ph{\lambda}\nu]\tau\rho\sigma} + \frac{1}{6}\nabla^{[\mu}R_{\lambda\tau\rho\sigma} J^{\nu]\lambda\tau\rho\sigma}+ \calO(S^4) \,. \label{eq:dsdt}
\end{align}
\end{subequations}
However, the introduction of the spin tensor adds 3 degrees of freedom to the problem. In order to close the system of equations describing the dynamics, we impose a spin supplementary condition (SSC). In particular we choose to impose the Tulczyjew-Dixon SSC~\cite{Tulc2,Dixon79} which reads
\begin{equation}\label{eq:ssc}
S^{\mu\nu}p_{\nu} = 0\,.
\end{equation}

Note that the mass $p_\mu p^\mu=-\tilde{m}^2$ is not conserved at $\calO(S^{2})$. Together with the equations of motion, the SSC allows us to find a conserved mass, that we use henceforth in this paper,
\begin{equation}\label{eq:defmass}
	m \equiv \tilde{m} - \frac{1}{6}R_{\mu\nu\rho\sigma}J^{\mu\nu\rho\sigma} \,,
\end{equation}
and to relate $p^{\mu}$ and the 4-velocity as
\begin{align}\label{eq:relationpu}
	p^{\mu} &= m u^{\mu} + \frac{1}{6} u^{\mu} R_{\rho\lambda\mu\nu}J^{\rho\lambda\mu\nu} - \frac{1}{2m} S^{\mu\nu}R_{\nu\lambda\rho\sigma}u^{\lambda}S^{\rho\sigma} + \frac{4}{3} R^{[\mu}_{\ph{\mu}\lambda\rho\sigma}J^{\nu]\lambda\rho\sigma} u_{\nu} \nn\\
	& \quad + \frac{2}{3} u_{\nu} \nabla^{\lambda}R^{[\mu}_{\ph{[\nu}\tau\rho\sigma}J_{\lambda}^{\ph{\lambda}\nu]\tau\rho\sigma} + \frac{1}{6} u_{\nu} \nabla^{[\mu}R_{\lambda\tau\rho\sigma}J^{\nu]\lambda\tau\rho\sigma} - \frac{1}{6m}S^{\mu\nu}J^{\lambda\tau\rho\sigma}\nabla_{\nu}R_{\lambda\tau\rho\sigma} + \calO(S^{4}) \,.
\end{align}
The norm of the spin tensor 
\begin{equation}
s^2\equiv \frac{1}{2}S_{\mu\nu}S^{\mu\nu},
\end{equation}
is conserved at the spin-cubic order that we consider: $\dd s/\dd \tau=0$.

Imposing that the quadrupole and octupole moments are purely spin-induced with the appropriate symmetries, we find
\begin{subequations}
\begin{align}
	J^{\mu\nu\rho\sigma} &= \frac{3 \kappa}{m}u^{[\mu}S^{\nu]\lambda}S_{\lambda}^{\ph{\lambda}[\rho}u^{\sigma]}\,, \\
	J^{\lambda\mu\nu\rho\sigma} &= \frac{\lambda}{4 m^{2}} \left[ \Theta^{\lambda[\mu}u^{\nu]}S^{\rho\sigma} + \Theta^{\lambda[\rho}u^{\sigma]}S^{\mu\nu} \right.\nn\\
	& \qquad \quad \; - \Theta^{\lambda[\mu}S^{\nu][\rho}u^{\sigma]} - \Theta^{\lambda[\rho}S^{\sigma][\mu}u^{\nu]} \nn\\
	&\qquad \quad \; \left. - S^{\lambda[\mu}\Theta^{\nu][\rho}u^{\sigma]} - S^{\lambda[\rho}\Theta^{\sigma][\mu}u^{\nu]} \right] \,,
\label{eq:Jdef}
\end{align}
\end{subequations}
where we used the notation $\Theta^{\mu\nu} = S^{\mu\lambda}S^{\nu}_{\phantom{\nu}\lambda}$, with quadrupolar and octupolar parameters $\kappa$, $\lambda$ normalized to unity for a Kerr black hole.

With the SSC~\eqref{eq:ssc}, one can construct a spin vector $S^\mu$ from the spin tensor $S^{\mu\nu}$, that also has a conserved norm by construction. They are linked using the orthogonal unit vector $p^\mu/\tilde{m}$ (the mass correction~\eqref{eq:defmass} is irrelevant at our PN order in spin terms) through the relations
\begin{subequations}\label{eq:defscovector}
\begin{align}
S^{\mu\nu} &= \epsilon^{\mu\nu\rho\sigma}\frac{p_{\rho}}{\tilde{m}} S_{\sigma}\,, \\
S^{\mu} &= - \frac{1}{2} \epsilon^{\mu\nu\rho\sigma}\frac{p_{\nu}}{\tilde{m}}S_{\rho\sigma} \,,
\end{align}
\end{subequations}
where $\epsilon^{\mu\nu\rho\sigma}$ is the Levi-Civita tensor. One can further construct a 3-dimensional spin vector with conserved Euclidean norm $\bm{S}$ (see Section II. C in Ref.~\cite{BFMP15}), which we use in our final results.

For the two-body problem, the equations of motion (EOM) of the individual bodies as well as the precession evolution of the spins have been derived in harmonic coordinates in Refs.~\cite{BMFB13,BFMP15,M15,LS15a,LS15b} and known at the considered PN order of the problem so that we do not need to solve them here. The precession equation in terms of the 3-dimensional spin vector reads $\dot{\bm{S}} = \bm{\Omega}\times \bm{S}$ where $\bm{\Omega}$ starts at $\calO(2)$. We recall that in our problem, we have rescaled the spin variable as in Eq.~\eqref{eq:rescale} to explicitly write the $c$ factors.

\subsection{Effective stress-energy tensor and source densities}\label{subsec:tmunu}

Equations~(2.23)-(2.25) of Ref.~\cite{M15} give the general form of the stress-energy tensor in terms of $p^\mu$, $S^{\mu\nu}$, $J^{\mu\nu\rho\sigma}$ and $J^{\lambda\mu\nu\rho\sigma}$. One can rewrite this expression as
\begin{equation}\label{eq:TwithCD}
T^{\mu \nu} = \sum_{A}\left[ U^{\mu \nu}_{A} \delta_{A} + \nabla_{\alpha}\left(U_{A}^{\mu \nu \alpha} \delta_{A} \right) + \nabla_{\alpha}\nabla_{\beta}\left(U_{A}^{\mu \nu \alpha \beta} \delta_{A} \right) + \nabla_{\alpha} \nabla_{\beta} \nabla_{\gamma}\left(U_{A}^{\mu \nu \alpha \beta \gamma} \delta_{A} \right)\right]\,,
\end{equation}
where $\delta_A \equiv \delta^{(3)}[\mathbf{x}-\bm{y}_A(t)]$ is the usual three-dimensional Dirac distribution. In our case, the expression of the $U$'s, up to the octupolar level, are given by
\begin{subequations}
\begin{align}
U^{\mu \nu}_{A} &= \dfrac{1}{u_{A}^{0} \sqrt{-g}} \left(p_{A}^{(\mu}u_{A}^{\nu)} + \dfrac{1}{3} R\indices{_A^(^\mu_\lambda_\rho_\sigma}J_{A}^{\nu) \lambda\rho \sigma} + \dfrac{1}{6} \nabla_{\lambda} R\indices{_A^(^\mu_\xi_\rho_\sigma}J_{A}^{\underline{\lambda}\nu) \xi \rho \sigma} + \dfrac{1}{12} \nabla^{(\mu}R_{A \xi \tau \rho \sigma}J_{A}^{\nu) \xi \tau \rho \sigma}\right), \\
U_{A}^{\mu \nu \alpha} &= \dfrac{1}{3 u_{A}^{0} \sqrt{-g}} \left(3 u_A^{(\mu}S_A^{\nu)\alpha} - \dfrac{1}{2} R\indices{_A^(^\mu_\xi_\lambda_\sigma}J_{A}^{\underline{\alpha}\nu)\xi \lambda \sigma} - R\indices{_A^(^\mu_\xi_\lambda_\sigma}J_{A}^{\nu)\alpha \xi \lambda \sigma} + R\indices{_A^\alpha_\xi_\lambda_\sigma}J_{A}^{(\mu \nu) \xi \lambda \sigma} \right),\\
U_{A}^{\mu \nu \alpha \beta} &=  -\dfrac{2}{3 u_{A}^{0} \sqrt{-g}} J_{A}^{\alpha (\mu \nu) \beta},\\
U_{A}^{\mu \nu \alpha \beta \gamma} &= -\dfrac{1}{3 u_{A}^{0} \sqrt{-g}} J_{A}^{\gamma \beta (\mu \nu) \alpha},
\end{align}
\end{subequations}
where $u_A^0=\dd t/\dd\tau_A$. The goal of this section is to decompose the spatial and temporal indices of the stress-energy tensor in order to make the factors in $c$ explicit. After restoring the $c$ powers in the expressions, we first express the covariant derivatives in terms of the partial derivatives and the Christoffel symbols, then separate spatial and temporal indices. The practical formulas to perform this decomposition are given in Sec. II. B. of Ref.~\cite{HFB20b}. Then, the source densities are defined by
\begin{equation}\label{eq:sigma}
\sigma = \frac{T^{00}+T^{ii}}{c^{2}}, \qquad \sigma_{i} = \frac{T^{0i}}{c}, \qquad \sigma_{ij} = T^{ij}.
\end{equation}
These quantities source the PN potentials as well as the source multipole moments. We provide in Eqs.~\eqref{eq:sigmaexplicit} the values of their spin contribution. As we will see later on, $\sigma$ is required up to $\calO(7)$, $\sigma_i$ to $\calO(6)$ and $\sigma_{ij}$ to $\calO(4)$.

\subsection{Source multipole moments}
\label{subsec:general def source multipole}

In the PN-MPM approach, a crucial step to compute the radiative moments is  the derivation of the so-called source multipole moments, which are expressed as integrals over the source. The mass and current source moments are given by~\cite{BlanchetLR}
\begin{subequations}\label{eq:defILJL}
\begin{align}
	I_L(t) &= \mathop{\mathrm{FP}}_{B=0}\,\int \ud^3\mathbf{x}\,\left(\frac{r}{r_0}\right)^B \int^1_{-1} \ud z\left\{\delta_\ell\,\hat{x}_L\,\Sigma -\frac{4(2\ell+1)}{c^2(\ell+1)(2\ell+3)} \,\delta_{\ell+1} \,\hat{x}_{iL} \,\Sigma_i^{(1)}\right.\nn\\
	&\qquad\quad \left. +\frac{2(2\ell+1)}{c^4(\ell+1)(\ell+2)(2\ell+5)} \,\delta_{\ell+2}\,\hat{x}_{ijL}\Sigma_{ij}^{(2)}\right\}(\mathbf{x},t+z\,r/c)\,,\\
	J_L(t) &= \mathop{\mathrm{FP}}_{B=0}\,\varepsilon_{ab<i_\ell} \int \ud^3 \mathbf{x}\,\left(\frac{r}{r_0}\right)^B \int^1_{-1} \ud z\left\{\vph{\frac{1}{1}} \delta_\ell\,\hat{x}_{L-1>a} \,\Sigma_b \right. \nn\\
	&\qquad\quad \left. -\frac{2\ell+1}{c^2(\ell+2)(2\ell+3)} \,\delta_{\ell+1}\,\hat{x}_{L-1>ac} \,\Sigma_{bc}^{(1)}\right\} (\mathbf{x},t+z\,r/c)\,,
\end{align}\end{subequations}
where we have $\delta_{\ell}(z) = a_{\ell}(1-z^{2})^{\ell}$ and $a_{\ell} = (2\ell+1)!!/2^{\ell+1}\ell!$ and
\begin{equation}\label{eq:Sigma}
\Sigma = \frac{\tau^{00}+\tau^{ii}}{c^{2}}, \qquad \Sigma_{i} = \frac{\tau^{0i}}{c}, \qquad \Sigma_{ij} = \tau^{ij}.
\end{equation}
The explicit expressions of the source moments in terms of the source densities and the PN metric are given in Eqs.~(4.7) of Ref.~\cite{BDEI05dr}. Furthermore, the computation of the general expressions of the source moments requires the introduction of a regularization. In this paper, we use the Hadamard \textit{partie finie} regularization, denoted by FP in which we introduce the scale constant $r_0$ \cite{BFreg}. The integral over $z$ can be easily integrated after performing a PN expansion
\begin{equation}\label{eq:intdeltal}
	\int^1_{-1} dz~ \delta_\ell(z) \,\Sigma(\mathbf{x},t+z\,r/c) = \sum_{k=0}^{+\infty}\,\frac{(2\ell+1)!!}{(2k)!!(2\ell+2k+1)!!}\,\left(\frac{r}{c}\right)^{2k}\!\Sigma^{(2k)}(\mathbf{x},t)\,.
\end{equation}
For the computation of the radiative moments at the order we aim at, we also need to compute another set of moments called gauge moments $(W_{L},X_L,Y_L,Z_{L})$. They are required to low order and admit similar definitions that can be found, e.g., in Eqs.~(125) of Ref.~\cite{BlanchetLR}.\\

From Eq.~\eqref{eq:taumunu}, we see for example that $\Sigma_{ij}=|g|\sigma_{ij} + \tfrac{c^4}{16\pi G}\Lambda_{ij}$. This means that the multipoles are sourced by the densities derived in the previous section, as well as the non-linearities of the gravitational field $\Lambda^{\mu\nu}$ through the PN metric. The relation between the source and radiative moments, given in Sec.~\ref{subsec:non-linear}, shows that the source moments are to be computed at the same orders as the ones displayed in Table \ref{tab:cycles}. This implies that, looking at \eqref{eq:defILJL}, we require $\Sigma$ up to $\calO(7)$, $\Sigma_i$ to $\calO(6)$ and $\Sigma_{ij}$ to $\calO(4)$, which is why the source densities were required to these orders.

\subsection{Post-Newtonian metric and potentials}
\label{subsec:summarymetric}

In the post-Newtonian framework, the metric is parametrized by a set of elementary retarded-type potentials that satisfy sourced wave equations. In this problem, we need the 3PN metric, which is given by \cite{BFeom}
\begin{subequations}\label{eq:metricg}
\begin{align} 
  g_{00} &=  -1 + \frac{2}{c^{2}}V - \frac{2}{c^{4}} V^{2} + \frac{8}{c^{6}}
  \left(\hat{X} + V_k V_k + \frac{V^{3}}{6}\right)\nonumber \\
  & \qquad +\frac{16}{c^8}\left(-\frac{V^4}{24} -V V_k V_k- V\hat{X}+2 \hat{R}_k V_k +2 \hat{T} \right) +\calO(10)\,,\\ 
  g_{0i} & = - \frac{4}{c^{3}}
  V_{i} - \frac{8}{c^{5}} \hat{R}_{i}-\frac{8}{c^7}\left( V^2 V_i +V_k \hat{W}_{ik} + 2 \hat{Y}_i  \right) + \calO(9)\,,\\ 
  g_{ij} & = \delta_{ij} \left[1 +
    \frac{2}{c^{2}}V + \frac{2}{c^{4}} V^{2} +\frac{8}{c^6}
  \left(\hat{X} + V_k V_k + \frac{V^{3}}{6}\right)\right] + 
  \frac{4}{c^{4}}\hat{W}_{ij}\nonumber \\
  & \qquad + \frac{8}{c^6}\left( -2 V_i V_j +V \hat{W}_{ij}+2 \hat{Z}_{ij} \right)+\calO(8) \,,
\end{align}
\end{subequations}\noindent
where the potentials are defined below. Note that an extension of this parametrization is known at 4PN together with the definition of the potentials required to this order. They are displayed in Appendix A of Ref.~\cite{MHLMFB20}.\\

However, in the present paper, we do not need to use the full metric. Instead, we need to know the source of the multipole moments to consistent order. It turns out that the only potentials that have a non-zero contribution to the spin effects in the source multipole moments are the following
\begin{subequations}\label{eq:defpotentials}
\begin{align}
  V & = \Box_{\mathcal{R}}^{-1}[-4 \pi G\, \sigma]\;,\label{V} \\ 
  V_{i} &= \Box_{\mathcal{R}}^{-1}[-4 \pi G\, \sigma_{i}]\,, \\ 
  \hat{W}_{ij} & =  \Box_{{\cal R}}^{-1}\left[-4 \pi G\, (\sigma_{ij} - \delta_{ij} \sigma_{kk}) - \partial_{i} V \partial_{j}V\right]\,, \label{Wij} \\
  \hat{R}_{i} & = \Box_{\mathcal{R}}^{-1}\bigg[-4 \pi G\, (V \sigma_{i} - V_{i} \sigma) - 2\partial_{k} V \partial_{i} V_{k} - \frac{3}{2} \partial_{t} V \partial_{i} V\bigg]\,, \\
  \hat{X} &= \Box_{\mathcal{R}}^{-1}\bigg[\vphantom{\frac{1}{2}} - 4 \pi G\, V \sigma_{ii} + \hat{W}_{ij}\partial_{ij} V + 2 V_{i} \partial_{t} \partial_{i} V + V \partial_{t}^{2}V+ \frac{3}{2}(\partial_{t} V)^{2} - 2\partial_{i} V_{j}\partial_{j} V_{i}\bigg] \,, \\ \hat{Z}_{ij} &=
\Box^{-1}_\mathcal{R}\bigg[- 4\pi G (\sigma_{ij}-\delta_{ij}
\sigma_{kk})V - 2\partial_{(i} V \partial_t V_{j)} + \partial_i V_k \partial_j V_k + \partial_k V_i \partial_k
V_j - 2 \partial_{(i} V_k \partial_k V_{j)} \nonumber \\ &\qquad~  -
\delta_{ij} \partial_k V_l (\partial_k V_l - \partial_l V_k) -
\frac{3}{4} \delta_{ij} (\partial_t V)^2\bigg]\,,\label{Zij}
\end{align}
\end{subequations}
where $\Box^{-1}_\mathcal{R}$ refers to the retarded flat d'Alembertian. More specifically, we are interested in the spin contributions to these potentials as their point-particle part are already known. As we can see, the potentials are sourced by the source densities derived in the previous section as well as other simpler potentials. 

To find the specific orders at which we require the potentials, one has to look at the order at which the potentials enter in $\Sigma$, $\Sigma_i$ and $\Sigma_{ij}$ including also the order of appearance in the source densities $\sigma$, $\sigma_i$ and $\sigma_{ij}$. The orders required for each interaction are displayed in Table \ref{tab:potentials}.

\begin{table*}[h]
\caption{
Orders at which the potentials are required for the spin contributions in the full gravitational waveform at 3.5PN. LO refers to the leading order of the contribution of the interaction while RO refers to the relative order required.
\label{tab:potentials}}
\begin{center}
\begin{tabularx}{0.55\textwidth}{c *{2}{Y} *{2}{Y}}
\toprule
 Potential & \multicolumn{2}{c}{SO} & \multicolumn{2}{c}{SS} \\
\cmidrule(lr){2-3} \cmidrule(lr){4-5}
   & LO & RO & LO & RO  \\
\midrule
  $V$              & 3 & 2 & 0 & 0 \\
  $V_{i}$          & 1 & 3 & 0 & 0 \\
  $\hat{W}_{ij}$   & 1 & 2 & - & - \\
  $\hat{R}_{i}$    & 1 & 1 & - & - \\
  $\hat{X}$        & 1 & 0 & - & - \\
  $\hat{Z}_{ij}$   & 1 & 0 & - & - \\
\bottomrule
\end{tabularx}
\end{center}
\end{table*}
Note that the SSS interaction does not appear in Table \ref{tab:potentials} because it will play no role in the expressions of the source moments. More details regarding this affirmation are given in  the following section. The regularization scheme used to compute the potentials is the Hadamard regularization \cite{BI04mult}. The methods used to compute the spin part of the potentials to these orders are detailed in, e.g., Sec. IV of Ref.~\cite{MBFB13}. An interesting feature of the SS interaction is that, as explained in Ref.~\cite{BFMP15}, the potentials $V$ and $V_i$ contain distributional terms which are crucial to take into account for the computation of the source multipole moments.


\subsection{Integration of the source multipole moments}
\label{subsec:integration source multipole}

This step in the overall project is the most technical one. However, no new methods were required and we used the same computational techniques as, in e.g. Refs.~\cite{MHLMFB20,MBFB13,BFMP15,BFP98}. Thus, we refer to these articles for more details.

As discussed above, the $\Sigma$'s are composed of the source densities and the non-linearities of the gravitational field whose dependency comes through the potentials. Thus, the multipole moments are integrals sourced by three types of terms: the compact terms, the non-compact terms and the surface terms. Note that the sources of the moments can be written in different ways. For example a term of the form $\int \ud^3\mathbf{x}\, \hat{x}_L \partial_k V \partial_k V$ can be re-written, dropping here the FP notation, as
\begin{equation}\label{eq:example}
\int \ud^3\mathbf{x} \, \hat{x}_L \partial_k V \partial_k V = -4\pi G\int \ud^3\mathbf{x} \, \hat{x}_L \sigma V + \frac{1}{2}\int \ud^3\mathbf{x} \, \hat{x}_L \Delta(V^2)+ \calO(2).
\end{equation}
We see that the left-hand side, which is a non-compact term, can be turned into a sum of a compact and a so-called surface term\footnote{The second term of the right-hand side is called a surface term because we only need to know the expansion in $r$ of $V^2$ when $r\rightarrow\infty$ to compute it.}. The multipole moments have been independently computed and double checked using different formulations for their sources. It is crucial to take into account the distributional derivatives induced in the left-hand side of \eqref{eq:example} in order to recover the value of the right-hand side.\\

At this stage, after integrating Eqs.~\eqref{eq:defILJL} for required $\ell$ to consistent order, we obtain the expressions of the source moments in a general frame for arbitrary orbits. In particular, we assumed neither the aligned-spin nor the quasi-circular orbits conditions. The lengthy expressions of the source moments are not displayed in this paper, however we recall here their leading PN order for the spin contributions~\cite{M15}
\begin{subequations}\label{eq:ILJLLO}
\begin{align}
	\left(I_L\right)_{\rm NS} &= m_{1}y_{1}^{<L>} + \exch + \calO(2) \,, \\
	\left(J_L\right)_{\rm NS} &= y_{1}^{a}v_{1}^{b}\varepsilon^{ab<i_\ell} y_{1}^{L-1>} + \exch + \calO(2) \,, \\
	\left(I_L\right)_{\rm SO} &= \frac{2\ell}{c^{3}(\ell+1)} \left[ \ell v_{1}^{a}S_{1}^{b}\varepsilon^{ab<i_\ell} y_{1}^{L-1>} - (\ell-1)y_{1}^{a}S_{1}^{b}\varepsilon^{ab<i_\ell} v_{1}^{i_{\ell-1}}y_{1}^{L-2>} \right] + \exch + \calO(5) \,, \\
	\left(J_L\right)_{\rm SO} &= \frac{\ell+1}{2 c} S_{1}^{<i_{\ell}}y_{1}^{L-1>} + \exch + \calO(3) \,, \\
	\left(I_L\right)_{\rm SS} &= -\frac{\ell(\ell-1)\kappa_{1}}{2 m_{1}c^{4}}  S_{1}^{<i_{\ell}}S_{1}^{i_{\ell-1}}y_{1}^{L-2>} + \exch + \calO(6) \,, \\
	\left(J_L\right)_{\rm SS} &= \frac{(\ell-1)\kappa_{1}}{2 m_{1}c^{4}} \left[ 2 v_{1}^{a}S_{1}^{b}\varepsilon^{ab<i_\ell} S_{1}^{i_{\ell-1}}y_{1}^{L-2>} - (\ell-2)y_{1}^{a}v_{1}^{b}\varepsilon^{ab<i_\ell} S_{1}^{i_{\ell-1}}S_{1}^{i_{\ell-2}}y_{1}^{L-3>} \right] \nn\\ &\quad+ \exch + \calO(6) \,, \\
	\left(I_L\right)_{\rm SSS} &= \frac{\ell(\ell-1)(\ell-2)\lambda_{1}}{3(\ell+1) m_{1}^{2}c^{7}} \left[ -\ell v_{1}^{a}S_{1}^{b}\varepsilon^{ab<i_\ell} S_{1}^{i_{\ell-1}} S_{1}^{i_{\ell-2}} y_{1}^{L-3>} \right. \nn\\ 
	&\qquad \qquad \left.+ (\ell-3)y_{1}^{a}S_{1}^{b}\varepsilon^{ab<i_\ell} v_{1}^{i_{\ell-1}} S_{1}^{i_{\ell-1}} S_{1}^{i_{\ell-2}} v_{1}^{i_{\ell-3}}y_{1}^{L-4>}  \right] + \exch + \calO(8) \,, \label{eq:ILSSS}\\
	\left(J_L\right)_{\rm SSS} &= -\frac{(\ell+1)(\ell-1)(\ell-2)\lambda_{1}}{12 m_{1}^{2}c^{5}} S_{1}^{<i_{\ell}} S_{1}^{i_{\ell-1}} S_{1}^{i_{\ell-2}} y_{1}^{L-3>} + \exch + \calO(6) \,,
\end{align}\end{subequations}
where NS refers to the non-spinning contributions. As stated in the previous section, the SSS interaction plays no role at the level of the radiative moments. Indeed, the only mass-type moment in which the SSS appears is for $\ell=2$ which vanishes due to the $\ell-2$ factor and similarly for the current quadrupole. The source current octupole has a non-zero contribution. However, to obtain the radiative current octupole, one has to perform a time derivative which operates on the spin vector. Since $\dot{\bm{S}}_A=\calO(2)$, the SSS contribution to the radiative moment is higher order.


\subsection{Reduction to quasi-circular orbits in the CM frame for aligned spins}
\label{subsec:CoMcirc}

In this section, we reduce the expressions of the source moments in the CM frame in the quasi-circular orbits approximation without precession. The first step is to express them in the CM frame. We define as usual the quantities $\tmass=m_1+m_2$, $\delta=(m_1-m_2)/\tmass$, $\nu=m_1 m_2/\tmass^2$, $r=|\bm{y}_1-\bm{y}_2|$, $\bm{n}=(\bm{y}_1-\bm{y}_2)/r$, $\bm{v}=\bm{v}_1-\bm{v}_2$, $\kappa_\pm=\kappa_1\pm\kappa_2$ and $\lambda_\pm=\lambda_1\pm\lambda_2$. The CM frame is defined as the frame in which the CM position of the system $G^i$ vanishes. It allows us to express the positions and velocities of the two compact objects in terms of the dynamical variables of the system
\begin{subequations}
\begin{align}
\bm{y}_1^\text{CM} &= \frac{m_2}{\tmass}r \bm{n} + \bm{z}\,,\\
\bm{y}_2^\text{CM} &= -\frac{m_1}{\tmass}r \bm{n} + \bm{z}\,,
\end{align}
\end{subequations}
and similarly for the velocities. The function $\bm{z}$ is a higher order quantity known from previous works \cite{BMFB13,BFMP15,M15}. We also introduce the following combinations of the individual spins
\begin{subequations}
\begin{align}
\bm{S}&=\bm{S}_1+\bm{S}_2\,,\\
\bm{\Sigma}&=\frac{\tmass}{m_2}\bm{S}_2-\frac{\tmass}{m_1}\bm{S}_1\,.
\end{align}
\end{subequations}
With these relations in hand, we obtain the source multipoles in the CM frame.\\

The next step is to impose the aligned-spins condition which drastically simplifies the computations. In particular, this implies that the orbital motion remains planar and that the two individual spins are aligned with the total angular momentum of the system. We can define the unitary vector $\bm{\ell}=\bm{n}\times\bm{v}/|\bm{n}\times\bm{v}|$. In the absence of precession, $\bm{\ell}$ is constant and coincides with $\bm{e}_Z$ defined in Sec.~\ref{subsec:notations}. The spin combinations $\bm{S}$ and $\bm{\Sigma}$ are also directed along $\bm{\ell}$ such that
\begin{subequations}
\begin{align}
\bm{S}=S_\ell \bm{\ell}\,,\\
\bm{\Sigma}=\Sigma_\ell \bm{\ell}\,.
\end{align}
\end{subequations}\\

The last step is the reduction to quasi-circular orbits. As for the non-spinning case, the acceleration $\bm{a}=\bm{a}_1-\bm{a}_2$ is directed along $\bm{n}$ as
\begin{equation}\label{eq:accCM}
\bm{a}= - r \omega^2 \bm{n}\,,
\end{equation}
which defines the orbital frequency $\omega$. Note that we neglect the radiation reaction force in the EOM because we are interested in the spin contributions to the acceleration. The spin contributions to the radiation reaction term is at least of order $\mathcal{O}(8)$, which is a higher order than required. The EOM \eqref{eq:accCM} are known from previous works \cite{BFMP15,M15}. In this approximation, $\bm{v}=-r\omega\bm{\lambda}$ where $\bm{\lambda}$ completes the time-dependent orthonormal basis $(\bm{n},\bm{\lambda},\bm{\ell})$ as $\bm{\lambda}=\bm{\ell}\times\bm{n}$. With this parametrization,
\begin{subequations}\label{eq:nlambda}
\begin{align}
\bm{n}(t)&=\cos\phi(t)\,\bm{e}_X+\sin\phi(t)\,\bm{e}_Y\,,\\
\bm{\lambda}(t)&= -\sin\phi(t)\,\bm{e}_X+\cos\phi(t)\,\bm{e}_Y \,,\\
\bm{\ell}(t)&= \bm{e}_Z\,,
\end{align}
\end{subequations}
where $\phi$ is the phase and is given by $\phi = \int \dd t\, \omega$. After defining the usual PN quantities
\begin{equation}
\gamma=\frac{G \tmass}{r c^2}\,, \qquad x=\left( \frac{G \tmass \omega}{c^3}\right)^{2/3},
\end{equation}
we can read the expression of $\omega^2$ in terms of $\gamma$. Then, we invert this relation to obtain $\gamma$ in terms of $\omega$ and thus $x$ which leads to
\begin{equation}
\gamma=x\left[ 1+ x g_\text{NS} + x^{3/2}\frac{g_\text{SO}}{G \tmass^2}+ x^2\frac{g_\text{SS}}{G^2 \tmass^4}+ x^{7/2}\frac{g_\text{SSS}}{G^3 \tmass^6}  + \calO(8) \right]\,,
\end{equation} 
where $g_\text{NS}$ and $g_\text{SO}$ are given in Eq.~(4.3) of Ref.~\cite{BMFB13}, $g_\text{SS}$ in Eq.~(3.32) of Ref.~\cite{BFMP15} and $g_\text{SSS}$ in Eq.~(6.15) of Ref.~\cite{M15} at consistent order. This allows us to express the source multipoles in terms of the orbital frequency and thus $x$.


\subsection{Non-linear contributions to the radiative moments}
\label{subsec:non-linear}

Once the source multipoles are known in the CM, we follow rigorously the procedure in Ref.~\cite{BFIS08} to compute the radiative moments. Without precession, there are no additional technicalities and the problem is equivalent to point-particle. 

To link the radiative to the source moments, one has to introduce the canonical moments  $(M_{L},S_{L})$ which are related to the set of source and gauge moments $(I_{L},J_{L},W_{L},X_{L},Y_{L},Z_{L})$ (see~\cite{BFIS08} for an account of the procedure). The computation of the canonical moments is detailed in Sec.~\ref{subsec:sourcecan}. Once the canonical moments are known, we can deduce the radiative moments through the following relations
\begin{subequations}
\begin{align}
U_L &= M_L^{(\ell)} + (\text{non-linear terms}),\\
V_L &= S_L^{(\ell)} + (\text{non-linear terms}),
\end{align}
\end{subequations}
where the non-linear terms are at least of order $\calO(3)$, so for radiative multipoles that are required at a low order, their contribution vanish. They are composed of three different types of contributions: the instantaneous terms, the tail terms and the memory terms. The computation of these different terms are detailed below.

\subsubsection{Link between source and canonical moments}\label{subsec:sourcecan}

The relations between the canonical and source moments take the form
\begin{subequations}
\begin{align}
	M_{L} &= I_{L} + \delta I_{L} \,,\\
	S_{L} &= J_{L} + \delta J_{L} \,,
\end{align}
\end{subequations}
where $\delta I_{L}$ and $\delta J_{L}$ are non-linear corrections made of products of source and gauge moments, and starting at the 2.5PN order. The full expressions of $\delta I_{L}$ and $\delta J_{L}$ up to the 3.5PN order are displayed in Sec. III. B. of Ref.~\cite{FBI15}. Recently, the expression for $\delta I_{ij}$ has been derived up to the 4PN order \cite{BFL22}. For the spin contributions, we have explicitly
\begin{subequations}
\begin{align}
	\left[ \delta I_{ij} \right]_{S} &= \calO(8) \,,\\
	\left[ \delta I_{ijk} \right]_{S} &= {12G\over c^5}\left[ I_{\langle ij}Y_{k\rangle }^{(1)} \right]_{S} +\calO(8) \,,\\
	\left[ \delta J_{ij}\right]_{S} &= {2G\over c^5}\left[\epsilon_{ab\langle i}\left(-I^{(3)}_{j\rangle b} W_a-2I_{j\rangle b}Y_{a}^{(2)} +I_{j\rangle b}^{(1)}Y_{a}^{(1)}\right)+3J_{\langle i}W_{j\rangle}^{(1)}\right]_{S} +\calO(8) \,,
\end{align}
\end{subequations}
where the spin parts of $W_i$, $Y_i$ and $J_i$ for aligned spins are given by $[J_i]_S =S_\ell\ell^i/c$, $[W_i]_S =\nu r \Sigma_\ell \lambda^i/4c$ and $[Y_i]_S =\nu r \omega \Sigma_\ell n^i/4c$. With these relations in hand, it is very straightforward to compute the canonical moments. They have been computed at the same orders as those displayed in Table \ref{tab:cycles}. After obtaining their expression, one has to perform a time differentiation to obtain the linear part of the radiative moments.

\subsubsection{Instantaneous terms}

At the 3.5PN order, the instantaneous terms are quadratic interactions of the canonical moments and their complete expressions are displayed in Sec. III.A.1. of Ref.~\cite{BFIS08}. The relevant spin contributions of these terms are the following
\begin{subequations}
\begin{align}
	\left[U_{ij}^{\rm inst} \right]_{S} &= \frac{2G}{c^{5}} \left[ \frac{1}{3} \varepsilon_{ab\langle i} M_{j\rangle a}^{(4)}S_{b}\right]_{S} + \calO(8) \,,\\
	\left[U_{ijk}^{\rm inst} \right]_{S} &= \frac{G}{c^{5}} \left[ {1\over5}\epsilon_{ab\langle  i}\left( -12S^{(2)}_{j\underline{a}}M^{(3)}_{k\rangle b} - 8M^{(2)}_{j\underline{a}}S^{(3)}_{k\rangle b} -3S^{(1)}_{j\underline{a}}M^{(4)}_{k\rangle b} -27M^{(1)}_{j\underline{a}}S^{(4)}_{k\rangle b}-S_{j\underline{a}}M^{(5)}_{k\rangle b} \phfrac\right.\right. \nn\\
    & \qquad\qquad\qquad \left.\left. -9M_{j\underline{a}}S^{(5)}_{k\rangle b} -{9\over4}S_{\underline{a}}M^{(5)}_{jk\rangle b}\right)+{12\over5}S_{\langle i}S^{(4)}_{jk\rangle} \right]_{S} + \calO(7) \,,\\
	\left[V_{ij}^{\rm inst} \right]_{S} &= {G\over7\,c^{5}} \left[ 4S^{(2)}_{a\langle i}M^{(3)}_{j\rangle a}+8M^{(2)}_{a\langle i}S^{(3)}_{j\rangle a} +17S^{(1)}_{a\langle i}M^{(4)}_{j\rangle a}-3M^{(1)}_{a\langle i}S^{(4)}_{j\rangle a}+9S_{a\langle i}M^{(5)}_{j\rangle a} \phfrac\right. \nn\\
    & \qquad\qquad \left. - 3M_{a\langle i}S^{(5)}_{j\rangle a}-{1\over4}S_{a}M^{(5)}_{ija}-7\epsilon_{ab\langle i}S_{\underline{a}}S^{(4)}_{j\rangle b} \right]_{S} + \calO(7) \,,\\
	\left[V_{ijk}^{\rm inst} \right]_{S} &= -\frac{2G}{c^{3}} \left[ \mathrm{S}_{\langle i}M^{(4)}_{jk\rangle } \right]_{S} + \calO(6) \,,\\
\left[V_{ijkl}^{\rm inst} \right]_{S} &= \frac{G}{c^3}\left[-\frac{35}{3}S^{(2)}_{\langle ij}M^{(3)}_{kl \rangle} -\frac{25}{3}M^{(2)}_{\langle ij}S^{(3)}_{kl \rangle}  -\frac{65}{6}S^{(1)}_{\langle ij}M^{(4)}_{kl \rangle}
-\frac{25}{6}M^{(1)}_{\langle ij}S^{(4)}_{kl \rangle} -\frac{19}{6}S_{\langle ij}M^{(5)}_{kl \rangle} \right. \nonumber\\
& \left. \qquad\qquad -\frac{11}{6}M_{\langle ij}S^{(5)}_{kl \rangle}-\frac{11}{12}S_{\langle i}M^{(5)}_{jkl \rangle}\right]_S+ \calO(5)\,.
\end{align}
\end{subequations}
They are treated as $\delta I_L$ and $\delta J_L$. In these equations, $S_i$ and $S_{ij}$ correspond to the current canonical moments and not the spin vector or tensor of the individual bodies. Note that these interactions are to be evaluated at the retarded time $T_R=T-R/c$.

\subsubsection{Tail terms}

The tails correspond to time integrals over the past of the source. Their expressions are known for each $\ell$. For the spin effects at the order considered, we only need to compute the following ones
\begin{subequations}\label{eq:tails}
\begin{align}
	\left[U_{ij}^{\rm tail} \right]_{S} &= \frac{2G \ADM}{c^3} \int^{+\infty}_{0} \ud \tau \left[ \ln \left(\frac{\tau}{2b}\right)+\frac{11}{12} \right] \left[ M^{(4)}_{ij} (T_R-\tau) \right]_{S} \,,\\
	\left[U_{ijk}^{\rm tail} \right]_{S} &= {2G \ADM\over c^3} \int^{+\infty}_{0} \ud\tau\left[ \ln \left(\frac{\tau}{2b}\right)+{97\over60} \right] \left[ M^{(5)}_{ijk} (T_R-\tau) \right]_{S} \,,\\
	\left[V_{ij}^{\rm tail} \right]_{S} &= {2G \ADM\over c^3} \int^{+\infty}_{0} \ud\tau \left[ \ln \left(\frac{\tau}{2b}\right)+{7\over6} \right] \left[ S^{(4)}_{ij} (T_R-\tau) \right]_{S} \,,\\
	\left[V_{ijk}^{\rm tail} \right]_{S} &= {2G \ADM\over c^3} \int^{+\infty}_{0} \ud\tau \left[ \ln \left(\frac{\tau}{2b}\right)+{5\over3} \right] \left[ S^{(5)}_{ijk} (T_R-\tau) \right]_{S} \,,\\
	\left[V_{ijkl}^{\rm tail} \right]_{S} &= {2G \ADM\over c^3} \int^{+\infty}_{0} \ud\tau \left[ \ln \left(\frac{\tau}{2b}\right)+{119\over60} \right] \left[ S^{(6)}_{ijkl} (T_R-\tau) \right]_{S} \,,
\end{align}
\end{subequations}
where we introduced an arbitrary time-scale constant $b$. The mass monopole $\ADM$, or ADM mass, differs from the total constant mass $\tmass$ through the relation $\ADM = \tmass + \bar{E}/c^{2}$ where $\bar{E}$ is the conservative binding energy of the system. This implies that the spin contributions in $\ADM$ start at $\calO(5)$ and thus do not need to be taken into account in the computation of the tail terms as they are of higher order.\\

To compute the tail integrals, we consider the aligned-spin case. There are no precession effects to consider here and the evolution of the dynamics of the binary is qualitatively the same as for the usual quasi-circular orbits, with the aligned conserved norm spins acting simply as constant vectors. This is to be contrasted with the more general case of binaries on quasi-circular but precessing orbits (as defined for instance in Ref.~\cite{ABFO08}), where one must solve analytically the dynamics consistently with the order at which the analysis is carried, to be able to compute these integrals. The idea to compute these integrals in the non-precessing case is to project the moments in the spatial basis $(\bm{e}_X,\bm{e}_Y,\bm{e}_Z)$ defined in Sec.~\ref{subsec:notations} using the relations \eqref{eq:nlambda} and assuming that the separation $r$ is constant over time\footnote{It has been shown \cite{BS93,ABIQ08} that for the tail integrals, such an approximation is valid because the remote past of the source is negligible when compared to its recent past. On this time scale, $r$ does not have the time to vary significantly and can be assumed constant.}. By doing so, we find one-dimensional integrals of the type $\int_0^\infty \dd y \ln(y) e^{\di \alpha y}$ that are computable analytically \cite{GR}.

\subsubsection{Memory terms}\label{subsubsec:memory}

The memory terms, as well as the tail terms, are called hereditary effects in the sense that they are integrals over the past of the source. They are integrals of quadratic interactions of canonical moments and only appear in the mass-type multipoles \cite{F09}. In our case, only the mass octupole moment contains a spin contribution to the memory terms according to 
\begin{equation}\label{eq:octmem}
\left[U_{ijk}^{\rm mem} \right]_{S} = -{4G\over 5 c^5} \int^{+\infty}_{0} \ud\tau \left[ \epsilon_{ab\langle i} \mathrm{M}^{(3)}_{j\underline{a}}(T_R-\tau)  \mathrm{S}^{(3)}_{k\rangle b} (T_R-\tau) \right]_{S}+\calO(7)\,.
\end{equation}
For the mass quadrupole, the spin contribution to the memory effects are of order $\calO(8)$ and thus do not need to be considered.

The computation of the memory terms differ from the one of the tails because one cannot assume that the separation $r$ is constant. We have to take into account the radiation reaction. As for the tails, we project the values of the moments on the basis $(\bm{e}_X,\bm{e}_Y,\bm{e}_Z)$. By doing so, we encounter integrals of the type
\begin{equation}
\int_{-\infty}^{T_R} \dd \tau\frac{e^{\di n \phi(\tau)}}{r^p(\tau)},
\end{equation}
where $n$ is a non-zero integer, $p$ is a half-integer, and $\phi$ is the phase variable defined in Sec.~\ref{subsec:CoMcirc}, which satisfies $\dot{\phi}=\omega$. As we can see, if we assumed the separation to be constant over time, this integral would diverge. At leading order, the separation scales as $r(\tau)\sim (-\tau)^{1/4}$ and the phase as $\phi(\tau)\sim (-\tau)^{5/8}$, which allows to compute this integral, as detailed in Refs.~\cite{ABIQ08,BFIS08}. 

One can also encounter integrals of this type with $n=0$, which can be computed analytically as well. However, the main difference from the case $n\neq 0$ is that they induce a factor $c^5$. This means that their contribution is of order 2.5PN lower than the initial order of the memory interaction. Thus, when using the PN-MPM formalism, we cannot be consistent to the 3.5PN order for these terms since it requires the knowledge of the general memory interactions at higher orders. Fortunately, these integrals only contribute to the modes for $m=0$, which we do not derive in this paper. More details about these peculiar modes are given in Sec. \ref{subsec:hlm}.\\

After combining all the previous intermediate results, we derived the radiative multipole moments in the quasi-circular, spin-aligned approximations at the orders displayed in Table \ref{tab:cycles}. To obtain the amplitude spherical modes, we insert the obtained expressions in Eq.~\eqref{eq:UlmVlm} and then in Eq.~\eqref{eq:hlm}.


\section{Results}\label{sec:results}

We now present the waveform modes. In Sec.~\ref{subsec:hlm}, we express them in a PN expansion following, e.g. \cite{BFIS08,FBI15}, whereas in Sec.~\ref{subsec:eob}, we factorize them in a way that is suitable for the EOB approach and notably template building.

\subsection{Spin-weighted spherical modes}\label{subsec:hlm}

The amplitude modes defined in Eq.~\eqref{eq:hlm} can be written in terms of the phase variable $\phi$ defined in Sec.~\ref{subsec:CoMcirc}. However it is convenient to introduce a new phase variable $\psi$ that allows factoring out the logarithm dependency on the orbital frequency induced by the tail terms in the radiative moments. The new phase variable reads
\begin{equation}\label{psi}
\psi \equiv \phi - \frac{2G \ADM \omega}{c^3} \ln\left(\frac{\omega}{\omega_0}\right),
\end{equation}
where the constant $\omega_0$ is linked to the time scale constant $b$ introduced in Eqs.~\eqref{eq:tails} through $\omega_0 = \tfrac{1}{4b}\text{exp}[\tfrac{11}{12}-\gamma_\text{E}]$ and we recall that $\ADM$ is the ADM mass. The amplitude modes then read
\begin{align}\label{eq:hlmexpl}
	h_{\ell m} = \frac{2 G \tmass \,\nu \,x}{R \,c^2} \,
	\sqrt{\frac{16\pi}{5}}\,\hat{H}_{\ell m}\,e^{-\di m \psi}\,,
\end{align}
where we recall that
\begin{equation}
x=\left(\frac{G \tmass\omega}{c^3}\right)^{2/3}\,.
\end{equation}
The spin part of the $\hat{H}_{\ell m}$ are given by
\begin{subequations}
\label{HlmExp}
\begin{align}
\hat{H}_{22}^\text{S} &= \frac{x^{3/2}}{G \tmass^2}\left[-2 S_{\ell} -  \tfrac{2}{3} \Sigma_{\ell} \delta + \bigl(S_{\ell} (- \tfrac{163}{63} -  \tfrac{92}{63} \nu) + \Sigma_{\ell} \delta(- \tfrac{1}{21}  + \tfrac{20}{63} \nu)\bigr) x \right. \nn\\
& \left. \qquad\qquad + \bigl((- \tfrac{4}{3}\di - 4 \pi) S_{\ell} -  \tfrac{4}{3} \pi \Sigma_{\ell} \delta\bigr) x^{3/2} \right. \nn\\
& \left. \qquad\qquad + \bigl(S_{\ell} (\tfrac{1061}{84} + \tfrac{4043}{84} \nu + \tfrac{499}{84} \nu^2) + \Sigma_{\ell}\delta (\tfrac{3931}{756} + \tfrac{7813}{378}  \nu + \tfrac{1025}{252}  \nu^2)\bigr) x^2\right]\nonumber \\
 & + \frac{x^2}{G^2 \tmass^4}\left[S_{\ell}^2 (2 + \kappa_+) + S_{\ell} \Sigma_{\ell} (2 \delta -  \kappa_- + \delta \kappa_+) + \Sigma_{\ell}^2 (- \tfrac{1}{2} \delta \kappa_- + \tfrac{1}{2} \kappa_+ - 2 \nu -  \kappa_+ \nu)\right. \nn\\
 & \left. \qquad\qquad+ \bigl(S_{\ell}^2 (- \tfrac{404}{63} + \tfrac{55}{42} \delta \kappa_- -  \tfrac{31}{42} \kappa_+ + \tfrac{68}{21} \nu + \tfrac{34}{21} \kappa_+ \nu)\right.\nn \\
& \left. \qquad\qquad  + S_{\ell} \Sigma_{\ell} (- \tfrac{481}{63} \delta + \tfrac{43}{21} \kappa_- -  \tfrac{43}{21} \delta \kappa_+ + \tfrac{68}{21} \delta \nu -  \tfrac{48}{7} \kappa_- \nu + \tfrac{34}{21} \delta \kappa_+ \nu) \right.\nn \\
& \left. \qquad\qquad + \Sigma_{\ell}^2 (- \tfrac{5}{3} + \tfrac{43}{42} \delta \kappa_- -  \tfrac{43}{42} \kappa_+ + \tfrac{172}{21} \nu -  \tfrac{89}{42} \delta \kappa_- \nu + \tfrac{25}{6} \kappa_+ \nu -  \tfrac{68}{21} \nu^2 -  \tfrac{34}{21} \kappa_+ \nu^2)\bigr) x \right.\nn \\
& \left. \qquad\qquad + \pi\bigl(S_{\ell}^2 (4 + 2 \kappa_+) + S_{\ell} \Sigma_{\ell} (4 \delta - 2 \kappa_- + 2 \delta \kappa_+) \right.\nn \\
& \left. \qquad\qquad\qquad + \Sigma_{\ell}^2 (- \delta \kappa_- + \kappa_+ - 4  \nu - 2  \kappa_+ \nu)\bigr) x^{3/2}\right]\nonumber \\
 & + \frac{x^{7/2}}{G^3 \tmass^6}\left[S_{\ell}^3 (\tfrac{32}{3} -  \tfrac{2}{3} \kappa_+ - 2 \lambda_+) + S_{\ell}^2 \Sigma_{\ell} (\tfrac{52}{3} \delta -  \tfrac{7}{3} \kappa_- -  \tfrac{1}{3} \delta \kappa_+ + 3 \lambda_- - 3 \delta \lambda_+) \right.\nn \\
& \left. \qquad\qquad + S_{\ell} \Sigma_{\ell}^2 (\tfrac{20}{3} - 3 \delta \kappa_- + 3 \kappa_+ + 3 \delta \lambda_- - 3 \lambda_+ -  \tfrac{112}{3} \nu -  \tfrac{2}{3} \kappa_+ \nu + 6 \lambda_+ \nu)\right. \\
& \left. \qquad\qquad  + \Sigma_{\ell}^3 (- \tfrac{5}{3} \kappa_- + \tfrac{5}{3} \delta \kappa_+ + \lambda_- -  \delta \lambda_+ -  \tfrac{20}{3} \delta \nu + \tfrac{11}{3} \kappa_- \nu -  \tfrac{1}{3} \delta \kappa_+ \nu - 3 \lambda_- \nu + \delta \lambda_+ \nu)\right]\,, \nonumber \\
\hat{H}_{21}^\text{S}  &= \frac{\di x}{2 G \tmass^2}\left[\Sigma_{\ell} + \bigl(- \tfrac{86}{21} S_{\ell} \delta + \Sigma_{\ell} (- \tfrac{79}{21} + \tfrac{139}{21} \nu)\bigr) x + \Sigma_{\ell} \bigl(- \tfrac{\di}{2} + \pi - 2\di \ln(2)\bigr) x^{3/2}\right.\nn \\
& \left. \qquad\qquad  + \bigl(S_{\ell}\delta (- \tfrac{331}{378}  + \tfrac{772}{189} \nu) + \Sigma_{\ell} (\tfrac{293}{378} -  \tfrac{2615}{756} \nu -  \tfrac{1723}{189} \nu^2)\bigr) x^2\right.\nn \\
& \left. \qquad\qquad  + \Bigl(S_{\ell}\delta \bigl(\tfrac{181}{105}\di  -  \tfrac{86}{21} \pi + \tfrac{172}{21}\di \ln(2)\bigr) + \Sigma_{\ell} \bigl(\tfrac{79}{42}\di -  \tfrac{79}{21} \pi -  \tfrac{1951}{140}\di \nu\right.\nn \\
& \left. \qquad\qquad\qquad  + \tfrac{257}{42} \pi \nu + \tfrac{158}{21}\di \ln(2) -  \tfrac{257}{21}\di \nu \ln(2)\bigr)\Bigr) x^{5/2}\right]\nonumber \\
 & + \frac{\di x^{5/2}}{G^2 \tmass^4}\left[S_{\ell}^2 (\delta -  \tfrac{1}{3} \kappa_- + \tfrac{1}{2} \delta \kappa_+) + S_{\ell} \Sigma_{\ell} (- \tfrac{1}{3} -  \tfrac{5}{6} \delta \kappa_- + \tfrac{5}{6} \kappa_+ - 4 \nu - 2 \kappa_+ \nu)\right.\nn \\
& \left. \qquad\qquad  + \Sigma_{\ell}^2 (- \tfrac{1}{2} \delta -  \tfrac{5}{12} \kappa_- + \tfrac{5}{12} \delta \kappa_+ -  \delta \nu + \tfrac{4}{3} \kappa_- \nu -  \tfrac{1}{2} \delta \kappa_+ \nu) \right.\nn \\
& \left. \qquad\qquad + \bigl(S_{\ell}^2 (\tfrac{41}{42} \delta + \tfrac{23}{48} \kappa_- + \tfrac{47}{336} \delta \kappa_+ -  \tfrac{1}{7} \delta \nu -  \tfrac{191}{72} \kappa_- \nu - \tfrac{1}{14} \delta \kappa_+ \nu)\right.\nn \\
& \left. \qquad\qquad  + S_{\ell} \Sigma_{\ell} (- \tfrac{29}{21} + \tfrac{19}{56} \delta \kappa_- -  \tfrac{19}{56} \kappa_+ + \tfrac{100}{21} \nu -  \tfrac{1301}{504} \delta \kappa_- \nu + \tfrac{1019}{504} \kappa_+ \nu + \tfrac{4}{7} \nu^2 +  \tfrac{2}{7} \kappa_+ \nu^2) \right.\nn \\
& \left. \qquad\qquad + \Sigma_{\ell}^2 (- \tfrac{6}{7} \delta + \tfrac{19}{112} \kappa_- -  \tfrac{19}{112} \delta \kappa_+ + \tfrac{59}{21} \delta \nu -  \tfrac{751}{504} \kappa_- \nu + \tfrac{145}{126} \delta \kappa_+ \nu + \tfrac{1}{7} \delta \nu^2 \right.\nn \\
& \left. \qquad\qquad \qquad+ \tfrac{1265}{504} \kappa_- \nu^2 + \tfrac{1}{14} \delta \kappa_+ \nu^2)\bigr) x\right] \\
 & + \frac{\di x^3}{2 G^3 \tmass^6}\left[S_{\ell}^2 \Sigma_{\ell} (1 + \tfrac{1}{2} \kappa_+) + S_{\ell} \Sigma_{\ell}^2 (\delta -  \tfrac{1}{2} \kappa_- + \tfrac{1}{2} \delta \kappa_+) + \Sigma_{\ell}^3 (- \tfrac{1}{4} \delta \kappa_- + \tfrac{1}{4} \kappa_+ -  \nu -  \tfrac{1}{2} \kappa_+ \nu)\right]\,, \nn \\
\hat{H}_{33}^\text{S}  &= \frac{3\di \sqrt{15} x^2}{8 \sqrt{14} G \tmass^2}\left[7 S_{\ell} \delta + \Sigma_{\ell} (3 - 9 \nu) + \bigl(S_{\ell} \delta(- \tfrac{139}{15}  + \tfrac{83}{15} \nu) + \Sigma_{\ell} (- \tfrac{43}{5} + 24 \nu + 5 \nu^2)\bigr) x \right.\nn \\
& \left. \qquad\qquad\qquad + \Bigl(S_{\ell}\delta \bigl(- \tfrac{213}{10}\di  + 21 \pi - 42\di  \ln(2) + 42\di  \ln(3)\bigr)\right.\nn \\
& \left. \qquad\qquad\qquad + \Sigma_{\ell} \bigl(- \tfrac{63}{5}\di + 9 \pi + \tfrac{8797}{270}\di \nu - 27 \pi \nu - 18\di \ln(2) + 54i \nu \ln(2)\right.\nn \\
& \left. \qquad\qquad\qquad\qquad + 18\di \ln(3) - 54\di \nu \ln(3)\bigr)\Bigr) x^{3/2}\right]\nonumber \\
 & + \frac{3\di \sqrt{15} x^{5/2}}{8 \sqrt{14} G^2 \tmass^4}\left[S_{\ell}^2\delta (-6  - 3 \kappa_+) + S_{\ell} \Sigma_{\ell} (-6 + 3 \delta \kappa_- - 3 \kappa_+ + 24 \nu + 12 \kappa_+ \nu)\right.\nn \\
& \left. \qquad\qquad\qquad + \Sigma_{\ell}^2 (\tfrac{3}{2} \kappa_- -  \tfrac{3}{2} \delta \kappa_+ + 6 \delta \nu - 6 \kappa_- \nu + 3 \delta \kappa_+ \nu)\right.\nn \\
& \left. \qquad\qquad\qquad + \bigl(S_{\ell}^2 (23 \delta -  \tfrac{7}{2} \kappa_- + \tfrac{15}{2} \delta \kappa_+ - 12 \delta \nu + 16 \kappa_- \nu - 6 \delta \kappa_+ \nu) \right.\nn \\
& \left. \qquad\qquad\qquad+ S_{\ell} \Sigma_{\ell} (26 - 11 \delta \kappa_- + 11 \kappa_+ - 128 \nu + 22 \delta \kappa_- \nu - 52 \kappa_+ \nu + 48 \nu^2 + 24 \kappa_+ \nu^2) \right.\nn \\
& \left. \qquad\qquad\qquad + \Sigma_{\ell}^2 (3 \delta -  \tfrac{11}{2} \kappa_- + \tfrac{11}{2} \delta \kappa_+ - 32 \delta \nu + \tfrac{59}{2} \kappa_- \nu -  \tfrac{37}{2} \delta \kappa_+ \nu + 12 \delta \nu^2\right.\nn \\
& \left. \qquad\qquad\qquad\qquad - 28 \kappa_- \nu^2 + 6 \delta \kappa_+ \nu^2)\bigr) x\right]\,, \\
\hat{H}_{32}^\text{S}  &= \frac{2 \sqrt{5} x^{3/2}}{3 \sqrt{7} G \tmass^2}\left[S_{\ell} + \Sigma_{\ell} \delta + \bigl(S_{\ell} (- \tfrac{13}{2} + \tfrac{73}{6} \nu) + \Sigma_{\ell}\delta (- \tfrac{31}{6}  + 5 \nu)\bigr) x\right.\nn \\
& \left. \qquad\qquad\qquad + \bigl( S_{\ell}(-\di + 2 \pi) + \Sigma_{\ell}\delta (-3\di  + 2 \pi)\bigr) x^{3/2}\right.\nn \\
& \left. \qquad\qquad\qquad + \bigl(S_{\ell} (\tfrac{4859}{1320} -  \tfrac{15413}{792} \nu -  \tfrac{419}{88} \nu^2) + \Sigma_{\ell}\delta (\tfrac{19241}{3960}  -  \tfrac{808}{55} \nu -  \tfrac{16153}{3960} \nu^2)\bigr) x^2\right]\nonumber \\
 & + \frac{8 \sqrt{5} x^3}{9 \sqrt{7} G^2 \tmass^4}\left[S_{\ell}^2 (-1 -  \tfrac{3}{8} \delta \kappa_- + \tfrac{9}{8} \kappa_+ -  \tfrac{9}{2} \nu -  \tfrac{9}{4} \kappa_+ \nu)\right.\nn \\
& \left. \qquad\qquad\qquad + S_{\ell} \Sigma_{\ell} (- \tfrac{5}{2} \delta -  \tfrac{3}{2} \kappa_- + \tfrac{3}{2} \delta \kappa_+ -  \tfrac{9}{2} \delta \nu + \tfrac{15}{4} \kappa_- \nu -  \tfrac{9}{4} \delta \kappa_+ \nu) \right.\nn \\
& \left. \qquad\qquad\qquad + \Sigma_{\ell}^2 (- \tfrac{3}{2} -  \tfrac{3}{4} \delta \kappa_- + \tfrac{3}{4} \kappa_+ + 3 \nu + \tfrac{3}{2} \delta \kappa_- \nu - 3 \kappa_+ \nu + \tfrac{9}{2} \nu^2 + \tfrac{9}{4} \kappa_+ \nu^2)\right]\nonumber \\
 & + \frac{4 \sqrt{5} x^{7/2}}{3 \sqrt{7} G^3 \tmass^6}\left[S_{\ell}^3 (1 + \tfrac{1}{2} \kappa_+) + S_{\ell}^2 \Sigma_{\ell} (2 \delta -  \tfrac{1}{2} \kappa_- + \delta \kappa_+)\right.\nn \\
& \left. \qquad\qquad\qquad + S_{\ell} \Sigma_{\ell}^2 (1 -  \tfrac{3}{4} \delta \kappa_- + \tfrac{3}{4} \kappa_+ - 5 \nu -  \tfrac{5}{2} \kappa_+ \nu)\right.\nn \\
& \left. \qquad\qquad\qquad + \Sigma_{\ell}^3 (- \tfrac{1}{4} \kappa_- + \tfrac{1}{4} \delta \kappa_+ -  \delta \nu + \kappa_- \nu -  \tfrac{1}{2} \delta \kappa_+ \nu)\right]\,, \\
\hat{H}_{31}^\text{S}  &= \frac{\di x^2}{24 \sqrt{14} G \tmass^2}\left[S_{\ell} \delta + \Sigma_{\ell} (5 - 15 \nu) + \bigl(S_{\ell}\delta (- \tfrac{79}{9}  + \tfrac{443}{9} \nu) + \Sigma_{\ell} (- \tfrac{149}{9} + \tfrac{700}{9} \nu -  \tfrac{841}{9} \nu^2)\bigr) x \right.\nn \\
& \left. \qquad\qquad\qquad + \Bigl(S_{\ell}\delta \bigl(\tfrac{47}{10}\di  + \pi - 2\di  \ln(2)\bigr) + \Sigma_{\ell} \bigl(-7\di + 5 \pi + \tfrac{11}{10}\di \nu - 15 \pi \nu\right.\nn \\
& \left. \qquad\qquad\qquad\qquad - 10\di \ln(2) + 30\di \nu \ln(2)\bigr)\Bigr) x^{3/2}\right]\nonumber \\
 & + \frac{\di x^{5/2}}{4 \sqrt{14} G^2 \tmass^4}\left[S_{\ell}^2 (\delta -  \tfrac{4}{3} \kappa_- + \tfrac{1}{2} \delta \kappa_+) + S_{\ell} \Sigma_{\ell} (1 -  \tfrac{11}{6} \delta \kappa_- + \tfrac{11}{6} \kappa_+ - 4 \nu - 2 \kappa_+ \nu)\right.\nn \\
& \left. \qquad\qquad\qquad + \Sigma_{\ell}^2 (- \tfrac{11}{12} \kappa_- + \tfrac{11}{12} \delta \kappa_+ -  \delta \nu + \tfrac{7}{3} \kappa_- \nu -  \tfrac{1}{2} \delta \kappa_+ \nu) \right.\nn \\
& \left. \qquad\qquad\qquad + \bigl(S_{\ell}^2 (- \tfrac{149}{18} \delta + \tfrac{13}{4} \kappa_- -  \tfrac{53}{36} \delta \kappa_+ -  \tfrac{22}{9} \delta \nu -  \tfrac{16}{9} \kappa_- \nu -  \tfrac{11}{9} \delta \kappa_+ \nu)\right.\nn \\
& \left. \qquad\qquad\qquad + S_{\ell} \Sigma_{\ell} (- \tfrac{115}{9} + \tfrac{85}{18} \delta \kappa_- -  \tfrac{85}{18} \kappa_+ + 40 \nu -  \tfrac{5}{9} \delta \kappa_- \nu + \tfrac{58}{9} \kappa_+ \nu + \tfrac{88}{9} \nu^2 + \tfrac{44}{9} \kappa_+ \nu^2)\right.\nn \\
& \left. \qquad\qquad\qquad + \Sigma_{\ell}^2 (- \tfrac{9}{2} \delta + \tfrac{85}{36} \kappa_- -  \tfrac{85}{36} \delta \kappa_+ + \tfrac{100}{9} \delta \nu -  \tfrac{233}{36} \kappa_- \nu + \tfrac{7}{4} \delta \kappa_+ \nu + \tfrac{22}{9} \delta \nu^2\right.\nn \\
& \left. \qquad\qquad\qquad\qquad -  \tfrac{2}{3} \kappa_- \nu^2 + \tfrac{11}{9} \delta \kappa_+ \nu^2)\bigr) x\right]\,, \\
\hat{H}_{44}^\text{S}  &= \frac{32 x^{5/2}}{9 \sqrt{35} G \tmass^2}\left[S_{\ell} (\tfrac{19}{3} - 19 \nu) + \Sigma_{\ell}\delta (3  - 6 \nu) \right.\nn \\
& \left. \qquad\qquad\qquad + \bigl(S_{\ell} (- \tfrac{437}{22} + \tfrac{10063}{132} \nu -  \tfrac{971}{44} \nu^2) + \Sigma_{\ell} \delta(- \tfrac{153}{11}  + \tfrac{2165}{66} \nu + \tfrac{67}{44} \nu^2)\bigr) x\right]\nonumber \\
 & + \frac{8 \sqrt{5} x^3}{9 \sqrt{7} G^2 \tmass^4}\left[S_{\ell}^2 (-4 - 2 \kappa_+ + 12 \nu + 6 \kappa_+ \nu)\right.\nn \\
& \left. \qquad\qquad\qquad + S_{\ell} \Sigma_{\ell} (-4 \delta + 2 \kappa_- - 2 \delta \kappa_+ + 12 \delta \nu - 6 \kappa_- \nu + 6 \delta \kappa_+ \nu)\right.\nn \\
& \left. \qquad\qquad\qquad + \Sigma_{\ell}^2 (\delta \kappa_- -  \kappa_+ + 4 \nu - 3 \delta \kappa_- \nu + 5 \kappa_+ \nu - 12 \nu^2 - 6 \kappa_+ \nu^2)\right]\,, \\
\hat{H}_{43}^\text{S}  &= \frac{9\di \sqrt{5} x^2}{8 \sqrt{14} G \tmass^2}\left[- S_{\ell} \delta + \Sigma_{\ell} (-1 + 3 \nu) + \bigl(S_{\ell}\delta (\tfrac{1303}{165}  -  \tfrac{1451}{165} \nu) + \Sigma_{\ell} (\tfrac{361}{55} -  \tfrac{284}{11} \nu + \tfrac{163}{11} \nu^2)\bigr) x\right.\nn \\
& \left. \qquad\qquad\qquad + \Bigl(S_{\ell}\delta \bigl(\tfrac{53}{10}\di  - 3 \pi + 6\di  \ln(2) - 6\di \ln(3)\bigr)\right.\nn \\
& \left. \qquad\qquad\qquad + \Sigma_{\ell} \bigl(\tfrac{32}{5}\di - 3 \pi -  \tfrac{6007}{270}\di \nu + 9 \pi \nu + 6\di \ln(2) - 18\di \nu \ln(2)\right.\nn \\
& \left. \qquad\qquad\qquad\qquad - 6\di \ln(3) + 18\di \nu \ln(3)\bigr)\Bigr) x^{3/2}\right]\nonumber \\
 & + \frac{9\di \sqrt{5} x^{7/2}}{8 \sqrt{14} G^2 \tmass^4}\left[S_{\ell}^2 (3 \delta + \tfrac{3}{5} \kappa_- -  \tfrac{8}{5} \delta \kappa_+ + 4 \delta \nu -  \tfrac{6}{5} \kappa_- \nu + 2 \delta \kappa_+ \nu) \right.\nn \\
& \left. \qquad\qquad\qquad + S_{\ell} \Sigma_{\ell} (6 + \tfrac{11}{5} \delta \kappa_- -  \tfrac{11}{5} \kappa_+ - 12 \nu -  \tfrac{16}{5} \delta \kappa_- \nu + \tfrac{48}{5} \kappa_+ \nu - 16 \nu^2 - 8 \kappa_+ \nu^2) \right.\nn \\
& \left. \qquad\qquad\qquad + \Sigma_{\ell}^2 (3 \delta + \tfrac{11}{10} \kappa_- -  \tfrac{11}{10} \delta \kappa_+ - 4 \delta \nu -  \tfrac{27}{5} \kappa_- \nu + \tfrac{16}{5} \delta \kappa_+ \nu - 4 \delta \nu^2 \right.\nn \\
& \left. \qquad\qquad\qquad\qquad + \tfrac{26}{5} \kappa_- \nu^2 - 2 \delta \kappa_+ \nu^2)\right]\,, \\
\hat{H}_{42}^\text{S}  &= \frac{4 x^{5/2}}{21 \sqrt{5} G \tmass^2}\left[S_{\ell} (- \tfrac{1}{9} + \tfrac{1}{3} \nu) + \Sigma_{\ell}\delta (1 - 2 \nu)\right.\nn \\
& \left. \qquad\qquad\qquad + \bigl(S_{\ell} (- \tfrac{43}{22} + \tfrac{6653}{396} \nu -  \tfrac{1387}{44} \nu^2) + \Sigma_{\ell} \delta (- \tfrac{313}{66} + \tfrac{3349}{198} \nu -  \tfrac{725}{44} \nu^2)\bigr) x\right]\nonumber \\
 & + \frac{4 x^3}{21 \sqrt{5} G^2 \tmass^4}\left[S_{\ell}^2 (\tfrac{5}{3} -  \tfrac{5}{4} \delta \kappa_- + \tfrac{25}{12} \kappa_+ - 5 \nu -  \tfrac{5}{2} \kappa_+ \nu) \right.\nn \\
& \left. \qquad\qquad\qquad + S_{\ell} \Sigma_{\ell} (\tfrac{5}{3} \delta -  \tfrac{10}{3} \kappa_- + \tfrac{10}{3} \delta \kappa_+ - 5 \delta \nu + \tfrac{15}{2} \kappa_- \nu -  \tfrac{5}{2} \delta \kappa_+ \nu) \right.\nn \\
& \left. \qquad\qquad\qquad + \Sigma_{\ell}^2 (- \tfrac{5}{3} \delta \kappa_- + \tfrac{5}{3} \kappa_+ -  \tfrac{5}{3} \nu + \tfrac{5}{2} \delta \kappa_- \nu -  \tfrac{35}{6} \kappa_+ \nu + 5 \nu^2 + \tfrac{5}{2} \kappa_+ \nu^2)\right]\,, \\
\hat{H}_{41}^\text{S}  &= \frac{\di \sqrt{5} x^2}{168 \sqrt{2} G \tmass^2}\left[S_{\ell} \delta + \Sigma_{\ell} (1 - 3 \nu) + \bigl(S_{\ell} \delta (- \tfrac{1147}{165} + \tfrac{1139}{165} \nu) + \Sigma_{\ell} (- \tfrac{309}{55} + \tfrac{232}{11} \nu -  \tfrac{111}{11} \nu^2)\bigr) x\right.\nn \\
& \left. \qquad\qquad\qquad + \Bigl(S_{\ell}\delta \bigl(- \tfrac{53}{30}\di  + \pi - 2\di  \ln(2)\bigr)\right.\nn \\
& \left. \qquad\qquad\qquad + \Sigma_{\ell} \bigl(- \tfrac{32}{15}\di + \pi + \tfrac{181}{30}\di \nu - 3 \pi \nu - 2\di \ln(2) + 6\di \nu \ln(2)\bigr)\Bigr) x^{3/2}\right]\nonumber \\
 & + \frac{\di \sqrt{5} x^{7/2}}{168 \sqrt{2} G^2 \tmass^4}\left[S_{\ell}^2 (-3 \delta -  \tfrac{7}{5} \kappa_- + \tfrac{12}{5} \delta \kappa_+ - 4 \delta \nu + \tfrac{14}{5} \kappa_- \nu - 2 \delta \kappa_+ \nu)\right.\nn \\
& \left. \qquad\qquad\qquad + S_{\ell} \Sigma_{\ell} (-6 -  \tfrac{19}{5} \delta \kappa_- + \tfrac{19}{5} \kappa_+ + 12 \nu + \tfrac{24}{5} \delta \kappa_- \nu -  \tfrac{72}{5} \kappa_+ \nu + 16 \nu^2 + 8 \kappa_+ \nu^2)\right.\nn \\
& \left. \qquad\qquad\qquad + \Sigma_{\ell}^2 (-3 \delta -  \tfrac{19}{10} \kappa_- + \tfrac{19}{10} \delta \kappa_+ + 4 \delta \nu + \tfrac{43}{5} \kappa_- \nu -  \tfrac{24}{5} \delta \kappa_+ \nu + 4 \delta \nu^2\right.\nn \\
& \left. \qquad\qquad\qquad\qquad -  \tfrac{34}{5} \kappa_- \nu^2 + 2 \delta \kappa_+ \nu^2)\right]\,, \\
\hat{H}_{55}^\text{S}  &= \frac{3125\di x^3}{36 \sqrt{66} G \tmass^2}\left[S_{\ell}\delta (- \tfrac{1}{2}  + \nu) + \Sigma_{\ell} (- \tfrac{1}{4} + \tfrac{5}{4} \nu -  \tfrac{5}{4} \nu^2)\right]\nonumber \\
 & + \frac{3125\di x^{7/2}}{96 \sqrt{66} G^2 \tmass^4}\left[S_{\ell}^2\delta (1 + \tfrac{1}{2} \kappa_+ - 2 \nu -  \kappa_+ \nu)\right.\nn \\
& \left. \qquad\qquad\qquad + S_{\ell} \Sigma_{\ell} (1 -  \tfrac{1}{2} \delta \kappa_- + \tfrac{1}{2} \kappa_+ - 6 \nu + \delta \kappa_- \nu - 3 \kappa_+ \nu + 8 \nu^2 + 4 \kappa_+ \nu^2)\right.\nn \\
& \left. \qquad\qquad\qquad + \Sigma_{\ell}^2 (- \tfrac{1}{4} \kappa_- + \tfrac{1}{4} \delta \kappa_+ -  \delta \nu + \tfrac{3}{2} \kappa_- \nu -  \delta \kappa_+ \nu + 2 \delta \nu^2 - 2 \kappa_- \nu^2 + \delta \kappa_+ \nu^2)\right]\,, \\
\hat{H}_{54}^\text{S}  &= \frac{32 x^{5/2}}{3 \sqrt{165} G \tmass^2}\left[S_{\ell} (-1 + 3 \nu) + \Sigma_{\ell} \delta(- 1 + 2 \nu)\right.\nn \\
& \left. \qquad\qquad\qquad + \bigl(S_{\ell} (\tfrac{241}{26} -  \tfrac{2939}{78} \nu + \tfrac{633}{26} \nu^2) + \Sigma_{\ell}\delta (\tfrac{619}{78}  -  \tfrac{300}{13} \nu + \tfrac{817}{78} \nu^2)\bigr) x\right]\,, \\
\hat{H}_{53}^\text{S}  &= \frac{3\di \sqrt{3} x^3}{4 \sqrt{110} G \tmass^2}\left[S_{\ell} \delta(\tfrac{1}{2}  - \nu) + \Sigma_{\ell} (- \tfrac{3}{4} + \tfrac{15}{4} \nu -  \tfrac{15}{4} \nu^2)\right]\nonumber \\
 & + \frac{9\di \sqrt{15} x^{7/2}}{32 \sqrt{22} G^2 \tmass^4}\left[S_{\ell}^2 (- \delta + \tfrac{4}{5} \kappa_- -  \tfrac{13}{10} \delta \kappa_+ + 2 \delta \nu -  \tfrac{8}{5} \kappa_- \nu + \delta \kappa_+ \nu)\right.\nn \\
& \left. \qquad\qquad\qquad + S_{\ell} \Sigma_{\ell} (-1 + \tfrac{21}{10} \delta \kappa_- -  \tfrac{21}{10} \kappa_+ + 6 \nu -  \tfrac{13}{5} \delta \kappa_- \nu + \tfrac{39}{5} \kappa_+ \nu - 8 \nu^2 - 4 \kappa_+ \nu^2)\right.\nn \\
& \left. \qquad\qquad\qquad + \Sigma_{\ell}^2 (\tfrac{21}{20} \kappa_- -  \tfrac{21}{20} \delta \kappa_+ + \delta \nu -  \tfrac{47}{10} \kappa_- \nu + \tfrac{13}{5} \delta \kappa_+ \nu - 2 \delta \nu^2 + \tfrac{18}{5} \kappa_- \nu^2 -  \delta \kappa_+ \nu^2)\right]\,, \\
\hat{H}_{52}^\text{S}  &= \frac{2 x^{5/2}}{9 \sqrt{55} G \tmass^2}\left[S_{\ell}(1 - 3 \nu)  + \Sigma_{\ell}\delta (1 - 2 \nu)  \right. \nonumber \\  
 & \left. \qquad \qquad \qquad  + ( S_{\ell}(- \tfrac{213}{26} + \tfrac{2519}{78} \nu -  \tfrac{493}{26} \nu^2) + \Sigma_{\ell}\delta(- \tfrac{535}{78} + \tfrac{244}{13} \nu -  \tfrac{565}{78} \nu^2)\bigr) x\right]\,, \\
\hat{H}_{51}^\text{S}  &= \frac{\di x^3}{216 \sqrt{385} G \tmass^2}\left[S_{\ell} \delta( 1 - 2\nu) + \Sigma_{\ell} (\tfrac{7}{2} -  \tfrac{35}{2} \nu + \tfrac{35}{2} \nu^2)\right]\nonumber \\
 & + \frac{\di \sqrt{5} x^{7/2}}{288 \sqrt{77} G^2 \tmass^4}\left[ S_{\ell}^2(\delta -  \tfrac{6}{5} \kappa_- + \tfrac{17}{10} \delta \kappa_+ - 2 \delta \nu + \tfrac{12}{5} \kappa_- \nu -  \delta \kappa_+ \nu) \right. \nonumber \\  
 & \left. \qquad \qquad \qquad \qquad + S_{\ell} \Sigma_{\ell}(1 -  \tfrac{29}{10} \delta \kappa_- + \tfrac{29}{10} \kappa_+ - 6 \nu + \tfrac{17}{5} \delta \kappa_- \nu -  \tfrac{51}{5} \kappa_+ \nu + 8 \nu^2 + 4 \kappa_+ \nu^2) \right. \nonumber \\  
 & \left. \qquad \qquad \qquad \qquad + \Sigma_{\ell}^2 (- \tfrac{29}{20} \kappa_- + \tfrac{29}{20} \delta \kappa_+ -  \delta \nu + \tfrac{63}{10} \kappa_- \nu -  \tfrac{17}{5} \delta \kappa_+ \nu + 2 \delta \nu^2 \right. \nonumber \\  
 & \left. \qquad \qquad \qquad \qquad\qquad -  \tfrac{22}{5} \kappa_- \nu^2 + \delta \kappa_+ \nu^2)\right]\,, \\
\hat{H}_{66}^\text{S}  &= -\frac{3132 x^{7/2}}{35 \sqrt{143} G \tmass^2}\left[ S_{\ell}(1 - 5 \nu + 5 \nu^2) +  \Sigma_{\ell}\delta (\tfrac{15}{29} - \tfrac{60}{29} \nu +  \tfrac{45}{29}\nu^2)\right]\,, \\
\hat{H}_{65} &= \frac{3125\di x^3}{144 \sqrt{429} G \tmass^2}\left[ S_{\ell}\delta(1- 2 \nu) + \Sigma_{\ell} (1 - 5 \nu + 5 \nu^2)\right]\,, \\
\hat{H}_{64}^\text{S}  &= \frac{256 \sqrt{2} x^{7/2}}{385 \sqrt{39} G \tmass^2}\left[S_{\ell} (1 - 5 \nu + 5 \nu^2) + \Sigma_{\ell} \delta (- \tfrac{5}{9}  + \tfrac{20}{9} \nu -  \tfrac{5}{3} \nu^2)\right]\,, \\
\hat{H}_{63}^\text{S}  &= -\frac{81\di x^3}{176 \sqrt{65} G \tmass^2}\left[S_{\ell} \delta (1 - 2 \nu) + \Sigma_{\ell} (1 - 5 \nu + 5 \nu^2)\right]\,, \\
\hat{H}_{62}^\text{S}  &= \frac{4 x^{7/2}}{693 \sqrt{65} G \tmass^2}\left[S_{\ell} (1 - 5 \nu + 5 \nu^2) + \Sigma_{\ell} \delta \left(\tfrac{17}{3} -  \tfrac{68}{3} \nu + 17  \nu^2 \right)\right]\,, \\
\hat{H}_{61}^\text{S}  &= \frac{\di x^3}{2376 \sqrt{26} G \tmass^2}\left[S_{\ell} \delta(1 - 2 \nu) +  \Sigma_{\ell}(1 - 5 \nu + 5 \nu^2)\right]\,, \\
\hat{H}_{77}^\text{S}  &= 0 \,,
\\\hat{H}_{76}^\text{S}  &= \frac{324 \sqrt{3} x^{7/2}}{35 \sqrt{143} G \tmass^2}\left[S_{\ell}(1 - 5 \nu + 5 \nu^2) +  \Sigma_{\ell}\delta(1 - 4 \nu + 3 \nu^2) \right]\,, \\
\hat{H}_{75}^\text{S}  &= 0\,, \\
\hat{H}_{74}^\text{S}  &= -\frac{512 \sqrt{2} x^{7/2}}{1365 \sqrt{33} G \tmass^2}\left[S_{\ell}(1 - 5 \nu + 5 \nu^2) +  \Sigma_{\ell}\delta(1 - 4 \nu + 3 \nu^2) \right]\,, \\
\hat{H}_{73}^\text{S}  &= 0\,, \\
\hat{H}_{72}^\text{S}  &= \frac{4 x^{7/2}}{3003 \sqrt{3} G \tmass^2}\left[ S_{\ell}(1 - 5 \nu + 5 \nu^2) + \Sigma_{\ell}\delta(1 - 4 \nu + 3 \nu^2) \right]\,, \\
\hat{H}_{71}^\text{S}  &= 0\,.
\end{align}
\end{subequations}
As stated in Sec.~\ref{subsubsec:memory}, we do not derive the $h_{\ell 0}$ because the PN-MPM formalism does not allow computing the memory terms to consistent orders. However, other methods were used in the literature to compute these non-oscillatory modes and their spin contribution are displayed to the 2PN order in Appendix A of Ref.~\cite{Mitman:2022kwt}.

For planar motions, the modes for negative $m$ are related to those for positive $m$ through $h_{\ell,-m}(\phi)=(-1)^{\ell+m}h_{\ell m}^*(\phi+\pi)$ which translates for the amplitude to $\hat{H}_{\ell, -m}=(-1)^\ell \hat{H}_{\ell m}^*$ where the star notation refers to the complex conjugate. The modes displayed above are in agreement with the literature, and notably with the modes derived for a test particle around a Kerr black hole~\cite{TMSS96,Pan11}.


\subsection{Effective-One-Body factorized modes}
\label{subsec:eob} 

Conveniently for EOB waveform models, we write the PN-expanded waveform given by Eqs.~\eqref{eq:hlmexpl}--\eqref{HlmExp} in a factorized, resummed form as \cite{Damour:2007xr,Damour:2007yf,Pan11,DIN09}
\begin{equation}
\label{hlmFact}
h_{\ell m}^\text{F} = h_{\ell m}^{(N,\epsilon_p)}\hat{S}_\text{eff}^{(\epsilon_p)}T_{\ell m}e^{\di \delta_{\ell m}}f_{\ell m}\,,
\end{equation}
where $\epsilon_p$ is the parity of $\ell+m$: $\epsilon_p = 0$ if $\ell+m$ is even, and $\epsilon_p = 1$ if $\ell+m$ is odd. 
The first term $h_{\ell m}^{(N,\epsilon_p)}$ is the leading (Newtonian) order waveform, which is known for any ($\ell,m$)~\cite{Th80,K07}, and its explicit expression is given in, e.g., Eq.~(3) of Ref.~\cite{Pan11}. 
Note that the convention for the definition of the waveform modes differs by a global minus sign between this paper and Ref.~\cite{Pan11}. However this difference only affects $h_{\ell m}^{(N,\epsilon_p)}$, while the other factors in Eq.~\eqref{hlmFact} are not altered. 

The (dimensionless) effective source term $\hat{S}_\text{eff}$ is given by either the effective energy $E_\text{eff}$ or the orbital angular momentum $p_\phi$, both expressed as functions of $v\equiv\left(\tmass \omega\right)^{1/3}=\sqrt{x}$, such that\footnote{In this section, we use units in which $c=G=1$ to simplify the notation.}
\begin{equation}
\hat{S}_\text{eff} = \left\{
        \begin{array}{ll}
            \frac{E_\text{eff}(v)}{\mu} , & \quad \ell + m \text{ even} \\
           v\, \frac{p_\phi(v)}{\mu\tmass}, & \quad \ell + m \text{ odd}
        \end{array}
    \right. ,
\end{equation}
where $\mu\equiv m_1m_2/\tmass$ is the reduced mass, and $E_\text{eff}$ is related to the total energy  $E$ via the EOB energy map $E = \tmass \sqrt{1+2\nu \left(E_\text{eff}/\mu - 1\right)}$.
This source term is motivated by the Regge-Wheeler-Zerilli equation~\cite{Regge:1957td,Zerilli:1970se}, whose source depends on the stress-energy tensor for a test body in a Schwarzschild background

The factor $T_{\ell m}$ resums the infinite number of ``leading logarithms'' entering the tail effects~\cite{Poisson:1993vp,Blanchet:1997jj,AF97}, and is given by
\begin{equation}
T_{\ell m} = \frac{\Gamma\left(\ell + 1 - 2 i \hat{k}\right)}{\Gamma (\ell + 1)} e^{\pi \hat{k}} e^{2i \hat{k} \ln (2m\omega r_0)},
\end{equation}
where $\Gamma(...)$ is the Euler gamma function, $\hat{k}\equiv m \omega E$, $\omega$ is the orbital frequency, and the constant $r_0$ takes the value $2\tmass/\sqrt{e}$ to give agreement with waveforms computed in the test-body limit~\cite{Pan11}.

The remaining part of the factorized modes is expressed as an amplitude $f_{\ell m}$ and a phase $\delta_{\ell m}$, which are computed such that the expansion of $h_{\ell m}^\text{F}$ agrees with the PN-expanded modes in Eq.~\eqref{eq:hlmexpl}.
To improve the agreement with numerical-relativity waveforms, $f_{\ell m}$ is further resummed as~\cite{DIN09,Pan11} $f_{\ell m} = (\rho_{\ell m})^\ell$ to reduce the magnitude of the 1PN non-spinning coefficient, which grows linearly with $\ell$.
For spinning binaries, the non-spinning and spin contributions are separated for the odd $m$ modes, such that
\begin{align}
f_{\ell m} = \left\{
        \begin{array}{ll}
           \rho_{\ell m}^\ell, & \quad m \text{ even} \\
           (\rho_{\ell m}^\text{NS})^\ell + f_{\ell m}^\text{S}, & \quad m \text{ odd}
        \end{array}
    \right. , 
\end{align}
where $\rho_{\ell m}^\text{NS}$ is the non-spinning part of $\rho_{\ell m}$, while $f_{\ell m}^\text{S}$ is the spin part of $f_{\ell m}$.

To simplify the expressions for the factorized modes, and to be consistent with the notation used in the literature, we introduce the dimensionless symmetric and antisymmetric spin parameters
\begin{subequations}
\begin{align}
\chi_S & \equiv \frac{1}{2} \left(\chi_1 + \chi_2\right) = \frac{S_\ell+\delta\Sigma_\ell}{2\nu G \tmass^2} \,,\\
\chi_A & \equiv \frac{1}{2} \left(\chi_1 - \chi_2\right) = -\frac{\delta S_\ell+(1-2\nu)\Sigma_\ell}{2\nu G \tmass^2},
\end{align}
\end{subequations}
and define the following combinations of the spin-multipole constants and spins:
\begin{subequations}
\begin{align}
\tilde{\kappa}_S &\equiv \frac{1}{2} \left[\chi _1^2 (\kappa_1 - 1)+\chi _2^2 (\kappa_2 - 1)\right], \\
\tilde{\kappa}_A &\equiv \frac{1}{2} \left[\chi _1^2 (\kappa_1 - 1) - \chi _2^2 (\kappa_2 - 1)\right], \\
\tilde{\lambda}_S &\equiv \frac{1}{2} \left[\chi _1^3 (\lambda_1 - 1)+\chi _2^3 (\lambda_2 - 1)\right], \\
\tilde{\lambda}_A &\equiv \frac{1}{2} \left[\chi _1^3 (\lambda_1 - 1) - \chi _2^3 (\lambda_2 - 1)\right],
\end{align}
\end{subequations}
which equal zero for black holes.

For the $(2,2)$ mode, we obtain
\begin{subequations}
\label{Fmode22}
\begin{align}
\rho_{22} &= 1 
+ v^2 \left(\tfrac{55}{84}\nu-\tfrac{43}{42}\right) 
+ v^3 \left[\left(\tfrac{2}{3}\nu-\tfrac{2}{3}\right) \chi _S-\tfrac{2}{3}\delta\chi _A\right] 
+ v^4 \Big[ 
\tfrac{19583}{42336} \nu^2-\tfrac{33025}{21168}\nu-\tfrac{20555}{10584}\nonumber\\
&\qquad
+ \left(\tfrac{1}{2}-2 \nu \right) \chi _A^2+\delta  \chi _A \chi _S+\tfrac{1}{2}\chi _S^2
+ \tfrac{1}{2}\delta  \kappa_A+\kappa_S \left(\tfrac{1}{2}-\nu \right)
\Big] \nonumber\\
&\quad
+ v^5 \left[\delta  \left(-\tfrac{19}{42}\nu-\tfrac{34}{21}\right) \chi _A+\left(\tfrac{209}{126}\nu^2+\tfrac{49}{18}\nu-\tfrac{34}{21}\right) \chi _S\right] \nonumber\\
&\quad
+ v^6 \Big[
\delta  \left(\tfrac{89}{126}-\tfrac{781}{252}\nu\right) \chi _A \chi _S 
+\left(-\tfrac{27}{14}\nu^2-\tfrac{457}{504}\nu+\tfrac{89}{252}\right) \chi _A^2
+\left(\tfrac{10}{9}\nu^2 -\tfrac{1817}{504}\nu +\tfrac{89}{252}\right) \chi _S^2 \nonumber\\
&\qquad
+ \left(\tfrac{67}{84}-\tfrac{139}{168} \nu\right) \delta\tilde{\kappa }_A+\left(-\tfrac{27}{28}\nu^2-\tfrac{407}{168}\nu+\tfrac{67}{84}\right) \tilde{\kappa }_S
\Big] \nonumber\\
&\quad
+ v^7 \Big[
\delta  \left(\tfrac{97865}{63504}\nu^2 +\tfrac{50140}{3969}\nu +\tfrac{18733}{15876}\right) \chi _A
+\left(\tfrac{50803}{63504}\nu^3 -\tfrac{245717}{63504}\nu^2 +\tfrac{74749 }{5292}\nu +\tfrac{18733}{15876}\right) \chi _S \nonumber\\
&\qquad
+ \delta \chi _A^3 \left(\tfrac{1}{3}-\tfrac{4}{3}\nu\right)+\delta  (2 \nu +1) \chi _A \chi _S^2+\left(-4 \nu ^2-3 \nu +1\right) \chi _A^2 \chi _S+\left(\nu +\tfrac{1}{3}\right) \chi _S^3 \nonumber\\
&\qquad
+ \tilde{\kappa}_S \left[\left(\tfrac{1}{3}\nu+\tfrac{4}{3}\right) \delta\chi _A+\left(-2 \nu ^2-\tfrac{14}{3}\nu +\tfrac{4}{3}\right) \chi _S\right]
+\tilde{\kappa }_A \left[\left(\tfrac{4}{3}-\tfrac{7}{3}\nu\right) \chi _A+\left(\tfrac{4}{3}-2   \nu \right) \delta\chi _S\right] \nonumber\\
&\qquad
+ (\nu -1 ) \delta\tilde{\lambda }_A+(3 \nu -1) \tilde{\lambda }_S
\Big], \\
\delta_{22} &=\tfrac{7 }{3}\omega E 
+ \big(\omega E\big)^2 \left[\left(\tfrac{8}{3}\nu-\tfrac{4}{3}\right) \chi _S-\tfrac{4}{3}\delta \chi _A\right],
\end{align}
\end{subequations}
where we only write the non-spinning part to the order needed for the 3.5PN spin contributions. 
The energy $E$ in $\delta_{\ell m}$ is replaced by the Hamiltonian in EOB waveform models.

For the $(2,1)$ mode, we obtain
\begin{subequations}
\label{Fmode21}
\begin{align}
\rho_{21}^\text{NS} &=1 
+\left(\tfrac{23}{84}\nu -\tfrac{59}{56}\right) v^2
+\left(\tfrac{617}{4704}\nu^2 -\tfrac{10993}{14112}\nu -\tfrac{47009}{56448}\right) v^4,
\\
f_{21}^\text{S} &= v \left(-\tfrac{3}{2}\tfrac{\chi _A}{\delta} - \tfrac{3}{2}\chi _S\right)
+ v^3 \Big[\left(\tfrac{131}{84}\nu +\tfrac{61}{12}\right) \tfrac{\chi _A}{\delta} +\left(\tfrac{79}{84}\nu +\tfrac{61}{12}\right) \chi _S\Big] \nonumber\\
&\quad
+ v^4 \left[(-2 \nu -3) \chi _A^2+ \left(\tfrac{21}{2}\nu -6\right) \frac{\chi _A \chi _S}{\delta}+\left(\tfrac{1}{2}\nu -3\right) \chi _S^2 
+\left(-\nu -\tfrac{1}{2}\right) \tilde{\kappa }_S-\frac{\tilde{\kappa }_A}{2 \delta } \right] \nonumber\\
&\quad
+ v^5 \Big[
\left(-\tfrac{703}{112}\nu^2 +\tfrac{8797}{1008}\nu -\tfrac{81}{16}\right) \frac{\chi _A}{\delta }+\left(\tfrac{613}{1008}\nu^2 +\tfrac{1709}{1008}\nu -\tfrac{81}{16}\right) \chi _S \nonumber\\
&\qquad
+ \left(\tfrac{3}{4}-3 \nu \right) \frac{\chi _A^3}{\delta }+ \left(\tfrac{9}{4}-6 \nu \right) \frac{\chi _A \chi _S^2}{\delta }+\left(\tfrac{9}{4}-3 \nu \right) \chi _A^2 \chi _S+\tfrac{3}{4} \chi _S^3 \nonumber\\
&\qquad
+ \tfrac{3}{4} \tilde{\kappa }_A \chi _A
+ \left(\tfrac{3}{4}-3 \nu \right) \frac{\tilde{\kappa }_A  \chi _S}{\delta }
+\left(\tfrac{3}{4}-\tfrac{3}{2}\nu\right) \frac{\tilde{\kappa }_S \chi _A}{\delta }
+\left(\tfrac{3}{4}-\tfrac{3}{2}\nu\right) \tilde{\kappa }_S \chi _S
\Big] \nonumber\\
&\quad
+ v^6 \Big[
\left(\tfrac{5}{7}\nu ^2 -\tfrac{9287}{1008}\nu +\tfrac{4163}{252}\right) \chi _A^2
+\left(\tfrac{139}{72} \nu ^2-\tfrac{2633 }{1008}\nu+\tfrac{4163}{252}\right) \chi _S^2 \nonumber\\
&\qquad
+ \left(\tfrac{9487}{504} \nu ^2 -\tfrac{1636}{21}\nu +\tfrac{4163}{126}\right) \frac{ \chi _A \chi _S}{\delta }
+ \left(\tfrac{473}{84} \nu^2+\tfrac{8}{21}\nu+\tfrac{1}{21}\right)  \frac{\tilde{\kappa }_A}{\delta } \nonumber\\
&\qquad
+\left(\tfrac{5}{14}\nu^2+\tfrac{10}{21}\nu+\tfrac{1}{21}\right) \tilde{\kappa }_S
\Big],
\\
\label{eq:delta21}\delta_{21} &= \tfrac{2}{3}\omega E -\tfrac{25}{2} \nu  v^5
+ \big(\omega E\big)^2 \left[\left(-\tfrac{69}{140}\nu -\tfrac{17}{35}\right) \frac{\chi _A}{\delta } +\left(-\tfrac{41}{28}\nu -\tfrac{17}{35}\right) \chi _S\right].
\end{align}
\end{subequations}
In the odd $m$ modes, the functions $f_{\ell m}$ and $\delta_{\ell m}$ depend on $1/\delta$, which diverges for equal masses. 
However, since the leading order of these modes is proportional to $\delta$, the PN-expanded modes do not diverge.
Thus, in EOB models, one needs to treat the equal-mass limit separately, as discussed in Appendix~A of Ref.~\cite{Cotesta:2018fcv}.

We note that the $\mathcal{O}(v^6 \chi^2 \nu^2)$ terms in the $(2,1)$ mode disagree with those used in the \texttt{SEOBNRv4HM} model~\cite{Cotesta:2018fcv}.\footnote{
The difference, for black holes, is given by
\begin{equation}
f_{21}^\text{S,(this paper)} - f_{21}^\text{S,(\texttt{SEOBNRv4HM})} = 
v^6 \nu ^2  \left(\tfrac{165}{112} \chi _A^2+\tfrac{87}{56}\frac{ \chi _A \chi _S}{\delta}+\tfrac{165}{112} \chi _S^2\right).
\end{equation}
}
Those terms were used in \texttt{SEOBNRv4HM} based on unpublished results by one of the authors of this paper. However, we checked that the SS contributions to the (2,1) mode should be given by Eq.~\eqref{Fmode21}. 

Interestingly, we also find a discrepancy with literature \cite{DIN09,Pan:2011gk} in the NS part of $\delta_{21}$, which was required to derive the spin terms in the factorized waveform. The difference is in the radiation reaction term $\mathcal{O}(\nu v^5)$ which, in these papers, has a coefficient $-493/42$. After investigation, we found out that this value came from a wrong expression for the (2,1) mode in Ref.~\cite{BFIS08}, which was later corrected in an erratum, but the factorized mode was never corrected. The coefficient -493/42 should read -25/2 as we see in Eq.~\eqref{eq:delta21}.

Explicit expressions for the other modes are given in Appendix~\ref{app:eob}. 
We also provide all expressions as a Mathematica file in the Supplemental Materials~\cite{SuppMaterial}.

\section{Summary}

We computed the spin contributions to the spherical-harmonics modes of the GW polarizations to the 3.5PN order, for non-precessing spins in quasi-circular orbits.
We used the PN-MPM formalism to tackle the computation of the radiative multipole moments, which were required to a higher multipolar order than what was known in the literature.
We also derived the spin contributions to the hereditary tail terms, as well as other non-linear interactions between the moments.

Our results include all spin terms, i.e., all SO, SS and SSS terms to that order. We wrote the waveform modes in two forms: in the conventional PN-expanded form, as well as a factorized form convenient for the EOB approach. 
The factorized modes we obtained are in agreement with Refs.~\cite{Pan11,Cotesta:2018fcv} except for the three terms $\mathcal{O}(v^6 \chi^2 \nu^2)$ in the (2,1) mode used in \texttt{SEOBNRv4HM}. As stated in Ref.~\cite{Cotesta:2018fcv}, these terms came from private communications and are now corrected. We also corrected a NS term in the quantity $\delta_{21}$.

The results derived in this paper can be useful in improving analytical waveform models. Interestingly, preliminary implementation of some of our results in an EOB waveform model showed a good improvement when compared to numerical relativity~\cite{PrivateCom}.

Future work will focus on relaxing the non-precessing approximation as well as the quasi-circular-orbits condition, which is important to improve waveform models for eccentric orbits and precessing spins.

\section*{Acknowledgments}
We are grateful to Luc Blanchet, Alejandro Bohé, Alessandra Buonanno, François Larrouturou, Guillaume Faye, Maarten van de Meent, Jan Steinhoff and Justin Vines for interesting discussions.


\appendix


\section{Source densities} \label{app:sourcedensities}

The explicit expressions of the spin part of source densities are given by
\begin{subequations}
\begin{align}\label{eq:sigmaexplicit}
\sigma^\text{SO} &=-\frac{4}{c^7\sqrt{-g}}\partial_t\left(\delta_1 S_1^{ab}v_1^a V^b \right)+ \frac{1}{\sqrt{-g}}\partial_i \left\{\delta_1\left[-\frac{2}{c^3} S_1^{ia}v_1^a +\frac{4}{c^5}\left( S_1^{ia} V^a -S_1^{ia} V v_1^a \right) \right. \right. \nn \\
& \left. \left.\qquad\qquad+\frac{4}{c^7} \left(2 \hat{R}^a S_1^{ia}+ S_1^{ia}(v_1^k V^k)v_1^a -2S_1^{ia} V^2 v_1^a+2 S_1^{ia} V V^a-S_1^{ia}v_1^b \hat{W}_{ab} \right) \right]   \right\} \nn \\
& + \delta_1 \left\{ \frac{2}{c^5} \left[S_1^{ab}v_1^a\partial^b V+2 S_1^{ab} \partial^a V^b \right] +\frac{4}{c^7} \left[ S_1^{ab}v_1^a \partial_t V^b +S_1^{bk}v_1^a v_1^b \partial_aV^k \right. \right. \nn \\
&\left.\left. + S_1^{ab} v_1^k \partial_b \hat{W}_{ka}-S_1^{ab}V^a\partial^bV-2 S_1^{ab}\partial^b\hat{R}^a-2S_1^{ab}v_1^aV \partial^b V+ 4 S_1^{ab} V \partial^b V^a\right]  \right\} \nn \\
&  + 1\leftrightarrow 2 +\calO(8),\\
\sigma^\text{SS} &=\frac{\kappa_1}{m_1 \sqrt{-g}}\partial_{kl}\left\{\delta_1\left[ \frac{S_1^{ka}S_1^{la}}{2c^4} + \frac{1}{c^6} \left(S_1^{ka}S_1^{la}\left(\frac{3}{4}v_1^2+\frac{3}{2}V \right) -\frac{1}{2} S_1^{ka}S_1^{lb}v_1^a v_1^b  \right.\right.\right. \nn \\
& \left.\left.\left. +\frac{1}{2}(S_1^{ab}S_1^{ab})v_1^k v_1^l + S_1^{ab}\left(-S_1^{lb}v_1^a v_1^k-S_1^{kb}v_1^a v_1^l \right)\right) \right]  \right\}\nn \\
& -\frac{\kappa_1}{2 m_1} S_1^{ba}S_1^{bi}\partial_{ia}V \delta_1 + \frac{\kappa_1}{m_1 c^6 \sqrt{-g}}\partial_k \left[\delta_1\left( S_1^{ab}S_1^{kb}\partial_a V-(S_1^{ab}S_1^{ab})\partial_k V \right)\right]\nn \\
&+ \frac{\kappa_1}{2 m_1 c^6 \sqrt{-g}}\partial_t^2(\delta_1 S_1^{ab}S_1^{ab})+ \frac{\kappa_1}{m_1 c^6 \sqrt{-g}}\partial_t\partial_k\left[\delta_1\left( -S_1^{ab}S_1^{kb}v_1^a+(S_1^{ab}S_1^{ab})v_1^k \right)\right]\nn \\
&  + 1\leftrightarrow 2 +\calO(8),\\
\sigma^\text{SSS} &=\frac{\lambda_1}{3m_1^2 c^7}v_1^m S_1^{mi}S_1^{jl}S_1^{kl}\partial_{ijk}\delta_1 + 1\leftrightarrow 2 +\calO(8),\\
\sigma_i^\text{SO} &=\frac{1}{\sqrt{-g}}\partial_k\left\{\delta_1\left[ \frac{S_1^{ik}}{2c} +\frac{S_1^{ak}v_1^a v_1^i}{2c^3}+\frac{2}{c^5}(S_1^{ka}v_1^i V^a-S_1^{ka} v_1^a V) \right]  \right\}\nn \\
& +\frac{1}{\sqrt{-g}}\partial_t\left\{\delta_1\left[ \frac{S_1^{ia}v_1^a}{2c^3} +\frac{2}{c^5}(S_1^{ia}v_1^a V-S_1^{ia} V^a) \right]  \right\} \nn\\
& + \delta_1 \left\{ -\frac{S_1^{ia}\partial^a V}{2c^3} +\frac{1}{c^5}\left[ -S_1^{ia}v_1^k \partial^kVv_1^a+\frac{1}{2}v_1^i S_1^{ab}v_1^a\partial^bV -\frac{1}{2}S_1^{ia}v_1^a\partial_tV+S_1^{ib}v_1^a\partial^aV^b \right. \right. \nn \\
& \left.\left.\qquad\qquad +S_1^{ia}V\partial^a V+S_1^{ib}v_1^a\partial^b V^a-S_1^{ab}v_1^a\partial^b V^i+S_1^{ab}v_1^a\partial^iV^b \right]\right\}+ 1\leftrightarrow 2 +\calO(8),\\
\sigma_i^\text{SS} &= \frac{\delta_1}{2m_1 c^6}\left[2 \kappa_1 S_1^{ab}S_1^{bj}\partial^{j(a}V v_1^{i)} -S_1^{ab}S_1^{ib}\partial_t\partial_aV+S_1^{bj}S_1^{ib}v_1^a\partial_{aj}V+2 S_1^{bj}S_1^{ia}\partial_{aj}V^b \right. \nn \\
& \left. - S_1^{aj}S_1^{ib}v_1^a\partial_{bj}V\right]+ \frac{\kappa_1}{m_1 c^6 \sqrt{-g}}\partial_t \left[ \delta_1 \left(\frac{1}{2}S_1^{ab}S_1^{ab}\partial^i V-S_1^{ab}S_1^{ib}\partial^a V  \right)  \right]\nn\\
& +\frac{\kappa_1}{c^4 m_1 \sqrt{-g}}\partial_{kl}\left\{ v_1^{[i}S_1^{k]a}S_1^{la}\delta_1+\frac{v_1^{[i}S_1^{k]a}}{c^2}\delta_1\left[ \frac{1}{2}S_1^{la}v_1^2+3 S_1^{la}V-S_1^{lb}v_1^a v_1^b+S_1^{ab}v_1^bv_1^l \right]  \right\}\nn\\
& -\frac{\kappa_1}{2 m_1 c^4\sqrt{-g}}\partial_t\partial_k\left\{S_1^{ia}S_1^{ka}\delta_1+\frac{\delta_1}{c^2}\left[ S_1^{ia}S_1^{ka}\left(\frac{v_1^2}{2}+3V\right) -2 v_1^{(i}S_1^{k)b}S_1^{ab}v_1^a- S_1^{ia}S_1^{kb}v_1^av_1^b\right]  \right\}\nn\\
& + \frac{\kappa_1}{m_1 c^6\sqrt{-g}}\partial_k\left\{\delta_1\left[ S_1^{ia}S_1^{ka}v_1^l\partial^lV+\frac{1}{2}S_1^{ia}S_1^{ka}\partial_tV+S_1^{ab}S_1^{kb}v_1^i\partial^aV -2S_1^{ab}S_1^{ib}v_1^k\partial^aV \right. \right. \nn \\
& \left.\left. -S_1^{ab}S_1^{kb}\partial^aV^i+S_1^{ab}S_1^{ib}\partial^aV^k-\frac{1}{2}S_1^{ab}S_1^{kb}v_1^a\partial^iV+\frac{1}{2}S_1^{ab}S_1^{ab}v_1^k\partial^iV+S_1^{ab}S_1^{kb}\partial^iV^a \right. \right. \nn \\
& \left.\left. +\frac{1}{2}S_1^{ab}S_1^{ib}v_1^a\partial^kV-\frac{1}{2}S_1^{ab}S_1^{ab}v_1^i\partial^kV-S_1^{ab}S_1^{ib}\partial^kV^a\right]  \right\}+ 1\leftrightarrow 2 +\calO(7),\\
\sigma_i^\text{SSS} &=\frac{\lambda_1}{12m_1^2 c^5}S_1^{ij}S_1^{km}S_1^{lm}\partial_{jkl}\delta_1 + 1\leftrightarrow 2 +\calO(7),\\
\sigma_{ij}^\text{SO} &=\frac{1}{c\sqrt{-g}}\partial_k\left(\delta_1 v_1^{(i}S_1^{j)k} \right)+\frac{1}{c^3\sqrt{-g}}\partial_t\left(\delta_1 v_1^{(i}S_1^{j)k}v_1^{k} \right) \nn \\
& +2\frac{\delta_1}{c^3}S_1^{k(i}\left( \partial_k V^{j)}+\partial^{j)}V v_1^k-\partial^{j)}V^k -\frac{1}{2}v_1^{j)}\partial_k V  \right)+1\leftrightarrow 2+\calO(5)\,,\\
\sigma_{ij}^\text{SS} &=\frac{\kappa_1}{m_1 c^4 \sqrt{-g}} \left\{ -\delta_1 \sqrt{-g} S_1^{ab} S_1^{a(i}\partial^{j)b}V  +\partial_t\partial_k \left[\delta_1\left( S_1^{ia}S_1^{ja}v_1^k+S_1^{ka}S_1^{a(i}v_1^{j)}\right)\right]  \right. \nn \\
&\left. +\frac{1}{2} \partial_t^2 \left(\delta_1 S_1^{ia}S_1^{ja}\right)-\partial_k \left[ \delta_1 \left( S_1^{ka}S_1^{a(i}\partial^{j)}V+\frac{1}{2}S_1^{ia}S_1^{ja}\partial^k V \right) \right] \nn \right. \\
&\left. +\frac{1}{2}\partial_{kl}\left[ \delta_1\left(S_1^{ka}S_1^{la}v_1^i v_1^j-2 v_1^{(i}S_1^{j)a}S_1^{ka} v_1^l +S_1^{ia}S_1^{ja}v_1^k v_1^l  \right) \right]\right\}+1\leftrightarrow 2+\calO(5)\,,\\
\sigma_{ij}^\text{SSS} &= \calO(5),
\end{align}
\end{subequations}
where $v_1^2=v_1^k v_1^k$.

\section{Explicit quantities of the factorized modes}\label{app:eob}
In this Appendix, we write the explicit expressions for the factorized modes (see Sec.~\ref{subsec:eob}). The $(2,2)$ mode is given by Eq.~\eqref{Fmode22}, the $(2,1)$ mode by Eq.~\eqref{Fmode21}, while the other subdominant modes read
\begin{subequations}
\begin{align}
\rho_{33}^\text{NS}&= 1+\left(\tfrac{2}{3}\nu-\tfrac{7}{6}\right) v^2,\\
f_{33}^\text{S}&= v^3 \left[\left(\tfrac{19}{2}\nu-2\right) \frac{\chi _A}{\delta }+\left(\tfrac{5}{2}\nu-2\right) \chi _S\right] 
+ v^4 \bigg[
\left(\tfrac{3}{2}-6 \nu \right) \chi _A^2+(3-12 \nu ) \frac{\chi _A \chi _S}{\delta }+\tfrac{3 }{2}\chi _S^2 \nonumber\\
&\qquad
+ \left(\tfrac{3}{2}-6 \nu \right)  \frac{\tilde{\kappa }_A}{\delta }+\left(\tfrac{3}{2}-3 \nu \right) \tilde{\kappa }_S 
\!\bigg]
+ v^5 \left[\left(\tfrac{407}{30}\nu ^2-\tfrac{593}{60}\nu+\tfrac{2}{3}\right) \frac{ \chi _A}{\delta }+\left(\tfrac{241}{30}\nu ^2 +\tfrac{11}{20}\nu +\tfrac{2}{3}\right) \chi _S\right] \nonumber\\
&\quad
+ v^6 \bigg[
\left(-12 \nu ^2+\tfrac{11}{2}\nu -\tfrac{7}{4}\right) \chi _A^2
+\left(44 \nu ^2-\nu -\tfrac{7}{2}\right) \frac{\chi _A \chi _S}{\delta }
+\left(6 \nu ^2-\tfrac{27}{2}\nu -\tfrac{7}{4}\right) \chi _S^2 \nonumber\\
&\qquad
+ \left(-12 \nu ^2+\tfrac{11}{2}\nu -\tfrac{7}{4}\right) \chi _A^2
+\left(44 \nu ^2-\nu -\tfrac{7}{2}\right)  \frac{\chi _A \chi _S}{\delta }
+\left(6 \nu ^2-\tfrac{27}{2}\nu -\tfrac{7}{4}\right) \chi _S^2
\bigg]
,\\
\delta_{33} &=\tfrac{13 }{10} \omega  E
+ \big(\omega E\big)^2 \left[\left(\tfrac{7339}{540}\nu -\tfrac{81}{20}\right) \frac{\chi _A}{\delta }+\left(\tfrac{593}{108}\nu -\tfrac{81}{20}\right) \chi _S\right],
\end{align}
\end{subequations}

\begin{subequations}
\begin{align}
\rho_{32} &= 1 + v \frac{4 \nu \chi _S}{3 (1-3 \nu )}
+ v^2 \left[
\frac{-\tfrac{32}{27}\nu ^2+\tfrac{223}{54}\nu-\tfrac{164}{135}}{1-3 \nu }
-\frac{16 \nu ^2 \chi _S^2}{9 (1-3 \nu )^2}
\right] \nonumber\\
&\quad
+ v^3 \left[
\left(\tfrac{13 }{9}\nu +\tfrac{2}{9}\right) \frac{\delta \chi _A}{1-3 \nu }
+\left(\tfrac{607}{81}\nu ^3 +\tfrac{503}{81}\nu ^2 -\tfrac{1478}{405}\nu +\tfrac{2}{9}\right)  \frac{\chi _S}{(1-3 \nu )^2}
+ \frac{320 \nu ^3 \chi _S^3}{81 (1-3 \nu )^3}
\right] \nonumber\\
&\quad
+ v^4 \bigg[
\frac{\tfrac{77141}{40095} \nu ^4 -\tfrac{508474 }{40095}\nu ^3 -\tfrac{945121 }{320760}\nu ^2 +\tfrac{1610009 \nu }{320760}-\tfrac{180566}{200475}}{(1-3 \nu )^2}
+ \left(4 \nu ^2-3 \nu +\tfrac{1}{3}\right) \frac{\chi _A^2}{1-3 \nu } \nonumber\\
&\qquad
+ \left(-\tfrac{50}{27}\nu ^2-\tfrac{88}{27}\nu+\tfrac{2}{3}\right) \frac{\delta \chi _A \chi _S}{(1-3 \nu )^2} 
+ \left(-\tfrac{2452}{243} \nu ^4 -\tfrac{1997}{243} \nu ^3 +\tfrac{1435}{243}\nu ^2 -\tfrac{43}{27}\nu +\tfrac{1}{3}\right)  \frac{\chi _S^2}{(1-3 \nu )^3}\nonumber\\
&\qquad
+ \left(\tfrac{1 }{3} -\tfrac{1}{3}  \nu \right)  \frac{\delta\tilde{\kappa }_A}{1-3 \nu }
+\left(2 \nu ^2-\nu +\tfrac{1}{3}\right)  \frac{\tilde{\kappa }_S}{1-3 \nu }
\bigg] \nonumber\\
&\quad
+ v^5 \bigg[
\left(-\tfrac{1184225 }{96228}\nu ^5 -\tfrac{40204523}{962280} \nu ^4 +\tfrac{101706029 }{962280}\nu ^3 -\tfrac{14103833 }{192456}\nu ^2 +\tfrac{20471053}{962280}\nu -\tfrac{2788}{1215}\right)  \frac{\chi _S}{(1-3 \nu )^3} \nonumber\\
&\qquad
+\left(\tfrac{889673}{106920}\nu ^3-\tfrac{75737}{5346} \nu ^2+\tfrac{376177 }{35640}\nu -\tfrac{2788}{1215}\right)  \frac{\delta  \chi _A}{(1-3 \nu )^2}
+ \left(\tfrac{608}{81} \nu ^3+\tfrac{736}{81} \nu ^2 -\tfrac{16  }{9}\nu\right)  \frac{\delta \chi _A \chi _S^2}{(1-3 \nu )^3} \nonumber\\
&\qquad
+\left(\tfrac{96176 }{2187}\nu ^5 +\tfrac{43528}{2187} \nu ^4-\tfrac{40232 }{2187}\nu ^3 +\tfrac{376 }{81}\nu ^2 -\tfrac{8 \nu }{9}\right)  \frac{\chi _S^3}{(1-3 \nu )^4} \nonumber\\
&\qquad
+\left(-\tfrac{32 }{3}\nu ^3+8 \nu ^2-\tfrac{8}{9}\nu\right)  \frac{\chi _A^2 \chi _S}{(1-3 \nu )^2}
+ \left(\tfrac{8}{9} \delta  \nu ^2-\tfrac{8 }{9}\delta  \nu \right)  \frac{\chi _S \tilde{\kappa }_A}{(1-3 \nu )^2}\nonumber\\
&\qquad
+\left(-\tfrac{16 }{3}\nu ^3 +\tfrac{8 }{3}\nu ^2 -\tfrac{8}{9}\nu\right)  \frac{\chi _S \tilde{\kappa }_S}{(1-3 \nu )^2}
\bigg], \\
\delta_{32} &= \left(\tfrac{11}{5}\nu+\tfrac{2}{3}\right)  \frac{\omega  E}{1-3 \nu }
+ v^4 \left[
\frac{4 \delta  \chi _A}{1-3 \nu }
+\left(\tfrac{36}{5} \nu ^2-20 \nu +4\right) \frac{\chi _S}{(1-3 \nu )^2}
\right] \nonumber\\
&\quad
+ v^5 \left[
\left(-\tfrac{144 }{5}\nu ^3 +80 \nu ^2-16 \nu \right)  \frac{\chi _S^2}{(1-3 \nu )^3}
-\frac{16 \delta  \nu  \chi _A \chi _S}{(1-3 \nu )^2}
\right],
\end{align}
\end{subequations}

\begin{subequations}
\begin{align}
\rho_{31}^\text{NS} &= 1 + v^2\left(-\tfrac{2}{9}\nu-\tfrac{13}{18}\right), \\
f_{31}^\text{S} &= v^3 \left[\left(\tfrac{11}{2}\nu -2\right) \frac{\chi _A}{\delta }+\left(\tfrac{13}{2}\nu-2\right) \chi _S\right] \nonumber\\
&\quad
+ v^4 \left[
\left(-6 \nu -\tfrac{5}{2}\right) \chi _A^2
+(4 \nu -5)  \frac{\chi _A \chi _S}{\delta }
-\tfrac{5}{2} \chi _S^2 
+\left(2 \nu -\tfrac{5}{2}\right) \frac{\tilde{\kappa }_A}{\delta }
+\left(-3 \nu -\tfrac{5}{2}\right) \tilde{\kappa }_S \right] \nonumber\\
&\quad
+ v^5 \left[\left(-\tfrac{931}{18}\nu ^2+\tfrac{25}{36}\nu+\tfrac{38}{9}\right) \frac{ \chi _A}{\delta }
+\left(-\tfrac{5}{2}\nu ^2-\tfrac{35}{12}\nu+\tfrac{38}{9}\right) \chi _S\right] \nonumber\\
&\quad
+ v^6 \bigg[
\left(\tfrac{44}{3}\nu ^2 +\tfrac{25}{6}\nu +\tfrac{43}{12}\right) \chi _A^2
+\left(\tfrac{452}{3} \nu ^2-\tfrac{83  }{3}\nu+\tfrac{43}{6}\right)  \frac{\chi _A \chi _S}{\delta }
+\left(22 \nu ^2-\tfrac{35}{2}\nu +\tfrac{43}{12}\right) \chi _S^2 \nonumber\\
&\qquad
+ \left(\tfrac{76 }{3}\nu ^2-\tfrac{47}{6}\nu+\tfrac{43}{12}\right) \frac{\tilde{\kappa }_A}{\delta }
+\left(\tfrac{22}{3} \nu ^2-\tfrac{2}{3}\nu+\tfrac{43}{12}\right) \tilde{\kappa }_S
\bigg],\\
\delta_{31} &= \tfrac{13 }{30} \omega  E
+ \big(\omega E\big)^2 \left[\left(-\tfrac{9 }{4}\nu+\tfrac{61}{20}\right) \frac{\chi _A}{\delta }+\left(\tfrac{77}{20}\nu+\tfrac{61}{20}\right) \chi _S\right],
\end{align}
\end{subequations}

\begin{align}
\rho_{44} &= 1 + \frac{v^2}{1-3 \nu }\left(-\tfrac{175}{88} \nu ^2+\tfrac{587}{132}\nu -\tfrac{269}{220}\right)
+ \frac{v^3}{1-3 \nu}\left[ \left(\tfrac{13}{5}\nu-\tfrac{2}{3}\right) \delta \chi _A+\left(-\tfrac{14}{5}\nu ^2 +\tfrac{41}{15}\nu -\tfrac{2}{3}\right) \chi _S\right] \nonumber\\
&\quad
+ v^4 \Big[\tfrac{1}{2}\delta  \tilde{\kappa }_A +\left(\tfrac{1}{2}-\nu \right) \tilde{\kappa }_S+\left(\tfrac{1}{2}-2 \nu \right) \chi _A^2+\delta  \chi _A \chi _S+\tfrac{1}{2}\chi _S^2\Big] \nonumber\\
&\quad
+ \frac{v^5}{(1-3 \nu )^2} \Big[
\delta  \left(\tfrac{597}{440} \nu ^3-\tfrac{1933}{220} \nu ^2+\tfrac{10679}{1650}\nu -\tfrac{69}{55}\right) \chi _A \nonumber\\
&\qquad
+ \left(\tfrac{591}{44} \nu ^4 +\tfrac{8539}{440} \nu ^3 -\tfrac{2673}{100} \nu ^2+\tfrac{16571}{1650}\nu-\tfrac{69}{55}\right) \chi _S
\Big], 
\end{align}

\begin{subequations}
\begin{align}
\rho_{43}^\text{NS} &= 1 + \frac{v^2}{1-2 \nu } \left(-\tfrac{10}{11} \nu ^2+\tfrac{547}{176}\nu-\tfrac{111}{88}\right),\\
f_{43}^\text{S} &= \frac{v}{1-2 \nu}\left(\tfrac{5}{2} \nu  \chi _S- \tfrac{5}{2}\nu\frac{\chi _A}{\delta}\right) 
+ \frac{v^3}{1-2\nu} \left[
\left(\tfrac{887}{44}\nu -\tfrac{3143}{132} \nu ^2\right) \frac{\chi _A}{\delta}
+ \left(-\tfrac{529}{132} \nu ^2-\tfrac{667}{44}\nu\right) \chi _S
\right]\nonumber\\
&\quad
+ \frac{v^4}{1-2\nu} \Big[
\left(12 \nu ^2-\tfrac{37}{3}\nu +\tfrac{3}{2}\right) \chi _A^2
+ \left(\tfrac{137}{6} \nu ^2-18 \nu +3\right) \frac{\chi _A \chi _S}{\delta} 
+ \left(\tfrac{35}{6} \nu ^2+\tfrac{1 }{3}\nu+\tfrac{3}{2}\right) \chi _S^2 \nonumber\\
&\qquad
+ \left(6 \nu ^2-9 \nu +\tfrac{3}{2}\right) \frac{\tilde{\kappa }_A}{\delta}
+ \left(6 \nu ^2-6 \nu +\tfrac{3}{2}\right) \tilde{\kappa }_S
\Big],\\
\delta_{43} &= \left(\tfrac{4961}{810}\nu+\tfrac{3}{5}\right) \frac{\omega  E}{1-2 \nu } \nonumber\\
&\quad
+ \frac{v^4}{(1-2 \nu )^2} \left[
\left(\tfrac{17999}{324} \nu ^2-\tfrac{2605}{108}\nu+\tfrac{11}{4}\right) \frac{\chi _A}{\delta}
+ \left(\tfrac{2569}{324} \nu ^2-\tfrac{2011}{108}\nu+\tfrac{11}{4}\right) \chi _S
\right],
\end{align}
\end{subequations}

\begin{align}
\rho_{42} &= 1 + \frac{v^2}{1-3 \nu }\left(-\tfrac{19}{88} \nu ^2+\tfrac{353}{132}\nu-\tfrac{191}{220}\right) 
+ \frac{v^3}{1-3\nu} \left[\delta  \left(\tfrac{7}{5}\nu -\tfrac{2}{3}\right) \chi _A + \left(-\tfrac{26}{5} \nu ^2+\tfrac{59}{15}\nu-\tfrac{2}{3}\right) \chi _S\right] \nonumber\\
&\quad
+ \frac{v^4}{1-3\nu} \left[
\left(6 \nu ^2-2 \nu +\tfrac{1}{2}\right) \chi _A^2+\delta  \chi _A \chi _S+\tfrac{1}{2}\chi _S^2
+ \tfrac{1}{2}\delta  \tilde{\kappa }_A +3 \nu ^2 \tilde{\kappa }_S-\nu  \tilde{\kappa }_S+\tfrac{1}{2}\tilde{\kappa }_S +\tfrac{1}{2}\chi _S^2
\right] \nonumber\\
&\quad
+ \frac{v^5}{(1-3 \nu )^2} \Big[
 \left(\tfrac{2463}{88} \nu ^3-\tfrac{4309}{220} \nu ^2+\tfrac{2977}{825}\nu -\tfrac{7}{110}\right) \delta \chi _A \nonumber\\
&\qquad
+\left(\tfrac{393}{220} \nu ^4 +\tfrac{6437}{440} \nu ^3-\tfrac{12443}{1100} \nu ^2+\tfrac{1873}{825}\nu-\tfrac{7}{110}\right) \chi _S
\Big],
\end{align}

\begin{subequations}
\begin{align}
\rho_{41}^\text{NS} &= 1 + \frac{v^2}{1-2 \nu }\left(-\tfrac{6}{11} \nu ^2+\tfrac{1385}{528}\nu -\tfrac{301}{264}\right), \\
f_{41}^\text{S} &= \frac{v}{1-2 \nu } \left(\tfrac{5}{2} \nu  \chi _S-\tfrac{5}{2}\nu\frac{\chi _A}{\delta }\right)
+ \frac{v^3}{1 -2\nu} \left[
\left(\tfrac{783}{44}\nu-\tfrac{2207}{132} \nu ^2\right) \frac{\chi _A}{\delta }+\left(-\tfrac{841}{132} \nu ^2-\tfrac{563}{44}\nu\right) \chi _S
\right] \nonumber\\
&\quad
+\frac{v^4}{1-2\nu} \Big[
\left(12 \nu ^2-\tfrac{37}{3}\nu +\tfrac{3}{2}\right) \chi _A^2+\left(\frac{41 \nu ^2}{6}-18 \nu +3\right) \frac{\chi _A \chi _S}{\delta }+\left(\tfrac{35}{6}+\tfrac{1 }{3}\nu+\tfrac{3}{2}\right) \chi _S^2 \nonumber\\
&\qquad
+\left(-2 \nu ^2-9 \nu +\tfrac{3}{2}\right)  \frac{\tilde{\kappa }_A}{\delta }
+\left(6 \nu ^2-6 \nu +\tfrac{3}{2}\right) \tilde{\kappa }_S
\Big], \\
\delta_{41} &= \left(\tfrac{507}{10}\nu+\tfrac{1}{5}\right) \frac{\omega  E}{1-2 \nu }
+ \frac{v^4}{(1-2 \nu )^2} \left[\left(\tfrac{533}{4} \nu ^2-\tfrac{55}{12}\nu +\tfrac{11}{12}\right) \frac{\chi _A}{\delta }+\left(-\tfrac{1511}{12} \nu ^2-\tfrac{11  }{4}\nu+\tfrac{11}{12}\right) \chi _S\right],
\end{align}
\end{subequations}

\begin{subequations}
\begin{align}
\rho_{55}^\text{NS} &= 1, \\
f_{55}^\text{S} &= \frac{v^3}{1-2\nu} \left[\left(-\tfrac{110}{3} \nu ^2+\tfrac{70}{3}\nu-\tfrac{10}{3}\right) \frac{\chi _A}{\delta }
+\left(-10 \nu ^2+10 \nu -\tfrac{10}{3}\right) \chi _S\right] \nonumber\\
&\quad
+ v^4 \left[\left(\tfrac{5}{2}-10 \nu \right) \chi _A^2+(5-20 \nu ) \frac{\chi _A \chi _S}{\delta }+\tfrac{5 }{2}\chi _S^2
+\left(\tfrac{5}{2}-10 \nu \right) \frac{\tilde{\kappa }_A}{\delta }+\left(\tfrac{5}{2}-5 \nu \right) \tilde{\kappa }_S\right],
\end{align}
\end{subequations}

\begin{align}
\rho_{54} &= 1 + \frac{3 \nu v  \left(-\delta  \chi _A-2 \nu  \chi _S+\chi _S\right)}{5 \left(5 \nu ^2-5 \nu +1\right)} \nonumber\\
&\quad
+ v^2 \bigg\lbrace
\frac{33320 \nu ^3-127610 \nu ^2+96019 \nu -17448}{13650 \left(5 \nu ^2-5 \nu +1\right)} \nonumber\\
&\qquad
-\frac{18 \nu ^2 \left[(1-4 \nu ) \chi _A^2+2 \delta  (2 \nu -1) \chi _A \chi _S+(1-2 \nu )^2 \chi _S^2\right]}{25 \left(5 \nu ^2-5 \nu +1\right)^2}
\bigg\rbrace \nonumber\\
&\quad
+ v^3 \bigg\lbrace
\frac{\delta  \left(-1268785 \nu ^4+2296805 \nu ^3-1386497 \nu ^2+268024 \nu -9100\right) \chi _A}{68250 \left(5 \nu ^2-5 \nu +1\right)^2} \nonumber\\
&\qquad
+\frac{\left(1690430 \nu ^5-1436855 \nu ^4-174699 \nu ^3+239045 \nu ^2+426 \nu -9100\right) \chi _S}{68250 \left(5 \nu ^2-5 \nu +1\right)^2} \nonumber\\
&\qquad
- \frac{162 \nu ^3}{125 \left(5 \nu ^2-5 \nu +1\right)^3} \Big[(1-4 \nu ) \delta\chi _A^3+3 (1-2 \nu )^2 \delta\chi _A \chi _S^2 \nonumber\\
&\qquad\qquad
+ (-24 \nu ^2+18 \nu -3) \chi _S \chi _A^2
-(1-2 \nu )^3 \chi _S^3\Big]
\bigg\rbrace,
\end{align}

\begin{subequations}
\begin{align}
\rho_{53}^\text{NS} &= 1, \\
f_{53}^\text{S} &= \frac{v^3}{1-2\nu} \left[\left(-\tfrac{62}{3} \nu ^2+18 \nu -\tfrac{10}{3}\right) \frac{\chi _A}{\delta }+\left(-\tfrac{46}{3} \nu ^2+\tfrac{46}{3}\nu-\tfrac{10}{3}\right) \chi _S\right] \nonumber\\
&\quad
+ \frac{v^4}{1-2\nu} \Big[
\left(20 \nu ^2-15 \nu +\tfrac{5}{2}\right) \chi _A^2
+\left(8 \nu ^2-30 \nu +5\right) \frac{ \chi _A \chi _S}{\delta }
+\left(\tfrac{5}{2}-5 \nu \right) \chi _S^2 \nonumber\\
&\qquad
+ \left(4 \nu ^2-15 \nu +\tfrac{5}{2}\right) \frac{\tilde{\kappa }_A}{\delta }
+ \left(10 \nu ^2-10 \nu +\tfrac{5}{2}\right) \tilde{\kappa }_S
\Big],
\end{align}
\end{subequations}

\begin{align}
\rho_{52} &= 1 + \frac{3 \nu v  \left(-\delta  \chi _A-2 \nu  \chi _S+\chi _S\right)}{5 \left(5 \nu ^2-5 \nu +1\right)} \nonumber\\
&\quad
+ v^2 \bigg\lbrace
\frac{21980 \nu ^3-104930 \nu ^2+84679 \nu -15828}{13650 \left(5 \nu ^2-5 \nu +1\right)} \nonumber\\
&\qquad
-\frac{18 \nu ^2 \left[(1-4 \nu ) \chi _A^2+2 \delta  (2 \nu -1) \chi _A \chi _S+(1-2 \nu )^2 \chi _S^2\right]}{25 \left(5 \nu ^2-5 \nu +1\right)^2}
\bigg\rbrace \nonumber\\
&\quad
+ v^3 \bigg\lbrace
\frac{\left(-963865 \nu ^4+1907465 \nu ^3-1213877 \nu ^2+243364 \nu -9100\right) \delta  \chi _A}{68250 \left(5 \nu ^2-5 \nu +1\right)^2} \nonumber\\
&\qquad
+\frac{\left(1859270 \nu ^5-2079455 \nu ^4+471681 \nu ^3+17105 \nu ^2+25086 \nu -9100\right) \chi _S}{68250 \left(5 \nu ^2-5 \nu +1\right)^2} \nonumber\\
&\qquad
- \frac{162 \nu ^3}{125 \left(5 \nu ^2-5 \nu +1\right)^3} \Big[(1-4 \nu ) \delta\chi _A^3+3 (1-2 \nu )^2 \delta\chi _A \chi _S^2 \nonumber\\
&\qquad\qquad
+ (-24 \nu ^2+18 \nu -3) \chi _S \chi _A^2
-(1-2 \nu )^3 \chi _S^3\Big]
\bigg\rbrace,
\end{align}

\begin{subequations}
\begin{align}
\rho_{51}^\text{NS} &=1, \\
f_{51}^\text{S} &= \frac{v^3}{1-2\nu} \left[\left(-\tfrac{38}{3} \nu ^2+\tfrac{46}{3}\nu-\tfrac{10}{3}\right) \frac{\chi _A}{\delta }+\left(-18 \nu ^2+18 \nu -\tfrac{10}{3}\right) \chi _S\right] \nonumber\\
&\quad
+\frac{v^4}{1-2\nu} \Big[
\left(20 \nu ^2-15 \nu +\tfrac{5}{2}\right) \chi _A^2
+ \left(-8 \nu ^2-30 \nu +5\right) \frac{\chi _A \chi _S}{\delta }
+\left(\tfrac{5}{2}-5 \nu \right) \chi _S^2 \nonumber\\
&\qquad
+\left(-4 \nu ^2-15 \nu +\tfrac{5}{2}\right)  \frac{\tilde{\kappa }_A}{\delta }+\left(10 \nu ^2-10 \nu +\tfrac{5}{2}\right) \tilde{\kappa }_S
\Big],
\end{align}
\end{subequations}

\begin{align}
\rho_{66} &= 1 + v^3 \frac{\delta  \left(-100 \nu ^2+85 \nu -14\right) \chi _A+\left(110 \nu ^3-150 \nu ^2+83 \nu -14\right) \chi _S}{21 \left(5 \nu ^2-5 \nu +1\right)},
\end{align}

\begin{subequations}
\begin{align}
\rho_{65}^\text{NS} &= 1,\\
f_{65}^\text{S} &= -\frac{7 \nu  v \left[(1-3 \nu ) \chi _A+\delta  (\nu -1) \chi _S\right]}{2 \delta  (\nu -1) (3 \nu -1)},
\end{align}
\end{subequations}

\begin{align}
\rho_{64} &= 1 + v^3 \frac{\delta  \left(-60 \nu ^2+65 \nu -14\right) \chi _A+\left(150 \nu ^3-230 \nu ^2+103 \nu -14\right) \chi _S}{21 \left(5 \nu ^2-5 \nu +1\right)},
\end{align}

\begin{subequations}
\begin{align}
\rho_{63}^\text{NS} &= 1,\\
f_{63}^\text{S} &= -\frac{7 \nu  v \left[(1-3 \nu ) \chi _A+\delta  (\nu -1) \chi _S\right]}{2 \delta  (\nu -1) (3 \nu -1)},
\end{align}
\end{subequations}

\begin{align}
\rho_{62} &= 1 + v^3\frac{\delta  \left(-36 \nu ^2+53 \nu -14\right) \chi _A+\left(174 \nu ^3-278 \nu ^2+115 \nu -14\right) \chi _S}{21 \left(5 \nu ^2-5 \nu +1\right)},
\end{align}

\begin{subequations}
\begin{align}
\rho_{61}^\text{NS} &= 1,\\
f_{61}^\text{S} &= -\frac{7 \nu  v \left[(1-3 \nu ) \chi _A+\delta  (\nu -1) \chi _S\right]}{2 \delta  (\nu -1) (3 \nu -1)},
\end{align}
\end{subequations}

\begin{equation}
\rho_{76} = 1 -\frac{4 \nu v \left[\delta  (2 \nu -1) \chi _A+\left(2 \nu ^2-4 \nu +1\right) \chi _S\right]}{7 \left(7 \nu ^3-14 \nu ^2+7 \nu -1\right)},
\end{equation}

\begin{equation}
\rho_{74} = 1 -\frac{4 \nu v \left[\delta  (2 \nu -1) \chi _A+\left(2 \nu ^2-4 \nu +1\right) \chi _S\right]}{7 \left(7 \nu ^3-14 \nu ^2+7 \nu -1\right)},
\end{equation}

\begin{equation}
\rho_{72} = 1 -\frac{4 \nu v \left[\delta  (2 \nu -1) \chi _A+\left(2 \nu ^2-4 \nu +1\right) \chi _S\right]}{7 \left(7 \nu ^3-14 \nu ^2+7 \nu -1\right)}.
\end{equation}


\bibliography{ListeRef_HMK22}

\end{document}